\documentclass[a4paper, 12pt, twoside]{book}

\usepackage[francais]{babel}
\usepackage{fancyhdr}
\usepackage[latin1]{inputenc}
\usepackage{epsfig}
\usepackage{here}
\usepackage{amsmath, amssymb, amsfonts} 
\usepackage{bbm} 

\input amssym.def
\input amssym.tex

\def\double{\Bbb}

\def\RR{{\double R}}
\def\CC{{\double C}}

\def\dd{{\cal D}}

\def\ff{{\cal F}}

\def\ll{{\cal L}}

\def\uu{{\cal U}}

\def\pa{{\partial}}
\def\Ds{{\,\hbox{${\rm D}\!\!\!\!/\,$}}}
\def\ds{{\,\hbox{$\partial\!\!\!/$}}}
\def\Df{{^{f}\!\Ds}}

\def\ef{{^{f}\!e}}

\def\omegaf{{^{f}\! \omega}}

\def\svphi{{\scriptscriptstyle \varphi}}
\def\dnx{{\sp d^{n}\!x }}
\def\id{{\mathbbm 1}} 

\def\beq{\begin{eqnarray}}
\def\eeq{\end{eqnarray}}
\def\bar{\begin{array}} 
\def\ear{\end{array}} 
\def\nnum{\nonumber}
\def\noi{\noindent}
\def\sp{\!\!\!\!\!\!}
\def\spp{\sp\sp}
\def\sppp{\spp\spp}

\begin{document}

\thispagestyle{empty}

\begin{center}

CENTRE DE PHYSIQUE TH\'EORIQUE \footnote{\, UMR 6207

- Unit\'e Mixte de Recherche du CNRS, des Universit\'es de
Provence,  de la M\'editerran\'ee et  du Sud Toulon-Var

- Laboratoire affili\'e \`a la FRUMAM - FR 2291\\}
\\ CNRS--Luminy, Case 907\\
13288
Marseille Cedex 9\\ FRANCE\\

\vspace{2cm}

{\Large\textbf{ Fluctuation of Dirac operator\\}}
\vspace{0.3cm} {\Large\textbf{ and equivalence principle}}
\\

\vspace{1.5cm}

{\large Mathieu MARCIANTE}

\vspace{2.5cm}

{\large\textbf{Abstract}}
\end{center}

General Relativity formulated with Noncommutative geometry allows one to obtain, via the fluctuation of Dirac operator, an exact equivalence principle: generation of curvature and torsion from flat space. The fluctuation method presented in this report is applied on two examples.

\vspace{1cm}

\vskip 2truecm

\noi Stage of D.E.A.\footnote{D.E.A. Physique Théorique, Physique Mathématique et Physique des Particules}\\
\noi June 2007\\

\begin{center}
{\large Directed by Pr. Thomas Schücker \\}
\end{center}

\newpage
\thispagestyle{empty} ${}$

\newpage
\thispagestyle{empty} {\bf\Large Remerciements}
\vspace{8cm}

Un stage de recherche n'est réellement profitable à un étudiant, encore naïf bien qu'assoiffé de connaissances,
que s'il est guidé et soutenu par des personnes sages et attentives à son épanouissement intellectuel. \\
Ainsi, je veux remercier {\bf Thomas Schücker} de m'avoir offert ce stage et toutes ses précieuses explications sur le
vaste domaine que recouvre ce sujet de physique théorique. Sa gentillesse et sa disponibilité m'ont été d'un grand réconfort. \\

Ce rapport de stage, plus que tous ceux que j'ai pu effectuer jusqu'à présent, est enfin pour moi l'occasion de
remercier l'ensemble des enseignants qui ont contribué à ma formation universitaire et qui ont su, de par la
passion qu'ils vouent à leur discipline, mettre en exergue la beauté de cette science.

\newpage
\thispagestyle{empty} ${}$

\newpage
\tableofcontents

\newpage
{\renewcommand{\thechapter}{}\renewcommand{\chaptername}{}
\addtocounter{chapter}{-1}
\chapter{Introduction}\markboth{\sl INTRODUCTION}{\sl INTRODUCTION}}
Since almost one century, one of the major challenges of Physics is the unification of the two fundamental theories that are General Relativity and Quantum Mechanics. Noncommutative Geometry offers a new mathematical way to reach this purpose, slightly guided by mathematical concepts governing modern physics:
\begin{itemize}
\item geometrical concepts, obtained from gauge theories using differential geometry.
\item noncommutative algebra concepts, inspired from Heisenberg viewpoint of Quantum Mechanics.
\end{itemize}

Provided with Noncommutative Geometry, whose basic elements are spectral triples and a finite set of axioms, it becomes possible to integrate inside a single mathematical formalism gravitational, electromagnetic, weak and strong interactions. Gauge fields of the three last interactions then become, inside this formalism, connections of a discrete space and are interpreted as pseudo-forces associated to the gravitational field.

One of the elements of this formalism is a self-adjoint operator $\dd$, unbounded, with compact resolvent and acting on a Hilbert space. Applied to high energy physics, this operator will be the Dirac operator: $\dd = \Ds$. It has now two important roles:
\begin{itemize}
\item it determines fermion dynamics via Dirac equation.
\item it encodes the space-time metric and thus become a dynamical object.
\end{itemize}
Indeed, in the same way that one can determine the electric or magnetic field by measuring the shift of the spectral lines of atoms immersed in these fields (Stark and Zeeman effects respectively), the Dirac operator spectra allows one to retrieve informations on the gravitational field via the distance formula.\\

The purpose of this report is to show that the Noncommutative Geometry, when a spectral triple is chosen equivalent to a Riemannian geometry allows, via the fluctuation of Dirac operator, the obtention of an exact equivalence principle: generation of curvature and torsion from flat space.

It is not the purpose of this report to introduce the Noncommutative Geometry, its application to the context of high energy physics being easily found \cite{SCH01}, \cite{SCH02}, \cite{KRA01}, \cite{STE01}. Hence, we will restrict its introduction to the elements taking part to the fluctuation method. \\
We begin this report by a short reminder on General Relativity, which will lead us to introduce the equivalence principle and to show the necessary formalism for applying the Dirac operator in curved space. We then introduce the fluctuation method in a simple case, and formalize it to the more general case. We finish by exemplifying  this fluctuation method on two examples using coordinate transformations. The first one takes place in a space of dimension 2, using the coordinate transformation cartesian/polar, the second one takes place in a space of dimension 3, using the coordinate transformation cartesian/spherical.

\newpage
\chapter{A look back on General Relativity}
General relativity is the theory that describes gravitational interactions. Space-time is interpreted as a pseudo-Riemaniann manifold $M$ on which test masses follow, in absence of other forces, geodesics of the manifold. The pseudo-metric $g(x)$, dynamical variable of the theory, plays the role of the gravitational potential and the connection $\Gamma(x)$ that of the gravitational field \cite{GetS01}.\\
This chapter makes use of some definitions and theorems which are, for clarity, gathered in the appendix. We consider a manifold $M$ of dimension 4 standing for the space-time, on which a metric $g$ is defined.\\

One of the foundations of General Relativity is the covariance concept. To design this theory, Einstein was guided by the idea that equations allowing one to describe physics have to conserve the same mathematical structure in every inertial frames (covariance under the Poincaré group: $SO(3,1)\ +\ translations\ $ from Special Relativity) and in every frames only subjected to gravitational forces: frames in free fall (covariance under the diffeomorphims of the manifold: Diff($M$)). The mathematical entities that possess such property are qualified as covariant.\\

Another major concept of the theory is the equivalence principle. Starting from a reference frame in which the observer measures a gravitational field, one can always bring him, by an appropriate transformation (by the action of an element of Diff($M$) onto the coordinate system of this observer) to a reference frame whose motion compensates the gravitational field and cancels its effect, which means that physics will then be properly described by Special Relativity \cite{BLA01}. This corresponds to the usual scene in which an observer in an elevator become weightless after the pull up cable has been (intentionally !) cut. \\

However, this equivalence principle is local (in the sense ``spatially restricted''). Indeed, when the gravitational field is not uniform, an observer making experiments in a free falling laboratory, a spatially extended place, will experience the gradient of the gravitationnal field, and will thus detect tide-generating forces. The equivalence principle is thus:
\begin{center}
    Every free falling reference frame is locally equivalent to an inertial frame.
\end{center}
The equivalence principle is only exact on the world line of a point particle ; this is the Fermi coordinate system. \\

\section{Einstein theory and Einstein-Cartan theory}
The formalism of General Relativity, such as designed by Einstein, does not allow to insert spinor objects. This is however required if one wants to take into account for elementary particles, which are in the biggest proportion fermions. It is Cartan who allowed to use spinor objects within the framework of General Relativity, by making use of non holonomic frames. \\

In General relativity, the space-time metric $g$ plays the role of the dynamical field giving birth to gravity. The first goal is thus to determine the differential equation ruling this field. However, the metric being is a bilinear symmetric form and a differential equation for this kind of object does not make sense. For that purpose, one can use a frame that allows to represent the metric by a symmetric $4\times 4$ matrix: the metric tensor (of rank 2), for which a system of differential equatiosn makes sense. However, the metric is globally defined over the manifold, and the use of a local frame to describe it brings an arbitrary into the equations: the choice of frame is the gauge of the theory. We will look closer on the difference between Special Relativity and General Relativity at this point : \\

\noi$\bullet$ Special Relativity:
\begin{itemize}
\item The flat space-time is describe by the Minkowski metric $\eta$ :\\
$\Longrightarrow\ g_{\mu\nu} = \eta_{\mu\nu}$ for every inertial observers, ($g_{\mu\nu}$ is thus fixed).
\item Relativity principle : two inertial observers describing the same physics, their respective reference frame $e_{a}$ and $\tilde e_{a}$ are deduced from each other by a Lorentz transformation $\Lambda_{_{\ll}}$ such that:
\vspace{-0.2cm}
\beq
\left(\Lambda_{_{\ll}}^{-1T}\,\eta\ \Lambda^{-1}_{_{\ll}}\right)_{ab} &=& \tilde\eta_{ab} \\
\nnum &\Longrightarrow&\ \ {\left(\Lambda_{_{\ll}}\right)^{\mu}}_{\nu} \in SO(3,1)
\eeq
\end{itemize}
$\bullet$ General Relativity (Einstein viewpoint) :
\begin{itemize}
\item The curved space-time is described by an arbitrary metric $g$ solution of the differential Einstein equation: \\
$\Longrightarrow\ \left(g(x)\right)_{\mu\nu}$ is different for every observer.
\item Equivalence principle : considering the frame $\beta_{\mu}(x)$ of an observer describing a curved space-time, one can always get the reference frame $\tilde\beta_{\mu}(x)$ of an observer describing a flat space-time, by an appropriate transformation ${\left(\Lambda_{\varphi}(x)\right)^{\mu}}_{\nu}$:
\beq
\sppp\sppp\forall \left(g(x)\right)_{\mu\nu}\ \exists {\left(\Lambda_{\varphi(x)}\right)^{\mu}}_{\nu}\ \textrm{s.t.
:}\ \ \
\left(\Lambda^{-T}_{\varphi}(x)g(x)\Lambda^{-1}_{\varphi}(x)\right)_{\mu\nu} &=& \left(\tilde\eta (x)\right)_{\mu\nu} \\
\nnum &\Longrightarrow&\ \ {\left(\Lambda_{\varphi}(x)\right)^{\mu}}_{\nu} \in {^{\uu}GL_{4}(\RR)}
\eeq
\end{itemize}
where $\varphi \in \textrm{Diff($M$)}$ ; ${\left(\Lambda_{\varphi}(x)\right)^{\mu}}_{\nu}$ is the jacobian of the diffeomorphism $\varphi$. \\
The coordinate system $\tilde x^{\mu}$ related to the metric tensor $\left(\tilde\eta (x)\right)_{\mu\nu}$ is the Fermi coordinate system. This metric tensor is equal to $\eta_{\mu\nu}$ only along a geodesic line. \\
\newpage
The problem of inserting spinors in the Einstein formalism of General Relativity comes from the fact that Dirac matrices are defined using a Clifford algebra, which requires to use an orthonormal frame, whereas the group $GL_{4}(\RR)$ does not possesses a spin cover. \\

\noi However, and what follows will lead us to the Cartan solution, there are different ways for characterizing a metric $g$ by using a frame:
\begin{itemize}
\item One can choose a frame $\beta_{\mu}(x)$ (16 degrees of freedom), then the matrix which represents the metric tensor $g_{\mu\nu}$ characterizes, with its 10 degrees of freedom, the metric in this frame. The frame is indifferent, so the arbitrary on its choice (the gauge) corresponds to the 16-parameters group ${^{\uu}GL_{4}(\RR)}$:
\begin{center}
    10 (metric tensor) + 16 (frame) - 16 (gauge group) = 10 independent parameters.\\
\end{center}
\item One can choose an orthonormal frame $e^{a}$ associated to this metric (16 degrees of freedom). The matrix which represents the metric tensor does not possesses any degrees of freedom as it is given by $\eta_{\mu\nu}$ in this frame. The frame being orthonormal, the arbitrary on its choice corresponds to the 6-parameters Lorentz group $SO(3,1)$:
\begin{center}
    0 (metric tensor) + 16 (frame) - 6 (gauge group) = 10 independent parameters.\\
\end{center}
\end{itemize}
\vspace{0.8cm}
Einstein uses the first characterization method but makes the implicit choice to use only frames deriving from a coordinates system : holonomic frames. This implicit choice comes from the fact that the covariance is realized under local diffeomorphisms $\varphi$ of the manifold. Hence, the arbitrary is no more related to the choice of frame but rather to the choice of a coordinate system, which represents only 4 independent degrees of freedom, and forms a subset of $^{\uu}GL_{4} = \left\{\Lambda(x)\right\}$ (where $\Lambda(x)$ possesses 16 degrees of freedom) and does not form a Lie group. Hence, General Relativity, as designed by Einstein, does not allow one to use spinor objects and does not account for a gauge theory.

\section{General Relativity as gauge theory}
Cartan's approach to General Relativity allows one to insert spinors in the theory and to treat General Relativity as a gauge theory.\\
Einstein required the covariance under diffeomorphisms of the manifold, Diff($M$) (i.e. under the coordinate transformations : $x^{\mu} = \varphi^{\mu}(\tilde x)$). These diffeomorphisms induce an arbitrary on the choice of frame restricted to holonomic (dual) frames: $dx^{\mu},\ d\tilde x^{\nu}$. The set of elements which encodes the transformation laws ${\left(\Lambda_{\varphi}(x)\right)^{\mu}}_{\nu} \in {^{\uu}GL_{4}(\RR)}$ (which are jacobians of the diffeomorphisms $\varphi$ : $dx^{\mu} = {\left(\Lambda_{\varphi}(x)\right)^{\mu}}_{\nu} d\tilde x^{\nu}$) does not forms a Lie group. \\
Cartan will extend this set of elements to the whole group $^{\uu}GL_{4}(\RR)$, by requiring a covariance, no more under the change of local maps of the manifold: Diff($M$), but under the change of arbitrary frames ; two arbitrary frames $\beta_{\mu}(x),\ \tilde \beta_{\mu}(\tilde x)$ being related by an element $\Lambda(x)$ belonging to $^{\uu}GL_{4}(\RR)$. \\
The set of permitted frames being now extended to non-holonomic frames, the Gram-Schmidt theorem ensures that we can obtain orthonormal frames, which will be written $e_{a}$ in the following. The link between an orthonormal (dual) frame $e^{a}$, that Cartan called {\it rep\`ere mobile}, and an holonomic (dual) frame $dx^{\mu}$ is made by using a vierbein ${e^{a}}_{\mu} \in {^{\uu}GL_{4}(\RR)}$:
\beq
    e^{a} = {e^{a}}_{\mu}\ dx^{\mu}
\eeq
We thus have the relation :
\beq
\left(g(x)\right)_{\mu\nu}\ dx^{\mu}\ dx^{\nu} = \left(g(x)\right)_{\mu\nu}\ {\left(e^{-1}(x)\right)^{\mu}}_{a}\ {\left(e^{-1}(x)\right)^{\nu}}_{b}\ e^{a}\ e^{b} = \eta_{ab}\ e^{a}\ e^{b}
\eeq
\beq
\label{jauge}\Longrightarrow\ \ \ \left(e^{-1T}(x)\ g(x)\ e^{-1}(x)\right)_{ab} = \eta_{ab}
\eeq
This will allow one to reduce the gauge group and to insert spinors in the theory. \\

Before this, we can make a little remark concerning the symmetric gauge. We can see that equation (\ref{jauge}) expressed as:
\beq
\left(g(x)\right)_{ab} = \left(e^{T}(x)\ \eta\ e(x)\right)_{ab}
\eeq
reduces in the euclidian case to $\left(g(x)\right)_{ab} = \left(e^{T}(x)\ e(x)\right)_{ab}$. Hence, using the symmetric gauge means expressing the vierbein ${\left(e(x)\right)^{a}}_{\mu}$ by a symmetric matrix. Equation (\ref{jauge}) then reduces to:
\beq
    \left(g(x)\right)_{ab} = \left(e(x)\ e(x)\right)_{ab} = \left(e^{2}(x)\right)_{ab}
\eeq
which makes explicit the fact that, using the symmetric gauge, one formally defines the vierbein by the square root of the metric.\\

We return now to the gauge group. To every holonomic (dual is now implicit) frame, $dx^{\mu}$ and $dx_{\varphi}^{\mu}$, linked by a diffeomorphism $\varphi \in \textrm{Diff($M$)}$ such that :
\beq
x_{\varphi}^{\mu} &=& \varphi^{\mu}(x) \\
\nnum dx_{\varphi}^{\mu} &=& {\left(\Lambda_{\varphi}(x)\right)^{\mu}}_{\nu}\ dx^{\nu}
\eeq
are related mobile frames $e^{a}$ et ${^{\varphi}\!e}^{a}$ linked by an element ${\left(\Lambda_{_{\ll}}(x)\right)^{a}}_{b}$ such that :
\beq
dx_{\varphi}^{\mu} &=& {\left(\Lambda_{\varphi}(x)\right)^{\mu}}_{\nu}\ dx^{\nu} = {\left(\Lambda_{\varphi}(x)\right)^{\mu}}_{\nu} {\left(e^{-1}(x)\right)^{\nu}}_{b}\ e^{b} \\
\nnum &=& {\left({^{\varphi}\!e}^{-1}(x_{\varphi})\right)^{\mu}}_{a}\ {^{\varphi}\!e}^{a} \\
\textrm{where,}\ \ \ {^{\varphi}\!e}^{a} &=& {\left({^{\varphi}\!e}(x_{\varphi})\right)^{a}}_{\mu}\ {\left(\Lambda_{\varphi}(x)\right)^{\mu}}_{\nu} {\left(e^{-1}(x)\right)^{\nu}}_{b}\ e^{b} = {\left(\Lambda_{_{\ll}}(x)\right)^{a}}_{b}\ e^{b}
\eeq
\beq
\textrm{thus}\ \ \ {\left(\Lambda_{_{\ll}}(x)\right)^{a}}_{b} &=& {\left({^{\varphi}\!e}(\varphi(x))\right)^{a}}_{\mu}\ {\left(\Lambda_{\varphi}(x)\right)^{\mu}}_{\nu} {\left(e^{-1}(x)\right)^{\nu}}_{b}
\eeq
The frames $e^{a}$ and ${^{\varphi}\!e}^{a}$ being orthonormal, this transformation preserves the Minkowski metric tensor $\eta_{\mu\nu}$, that is by definition the property of the element $\Lambda_{_{\ll}}$ from the Lorentz group $SO(3,1)$. However, the 16 parameters ${\left(\Lambda_{_{\ll}}\right)^{\mu}}_{\nu}$ are constant parameters whereas parameters ${\left(\Lambda_{_{\ll}}(x)\right)^{a}}_{b}$ are functions on $M$:
\beq
    {\left(\Lambda_{_{\ll}}(x)\right)^{a}}_{b} \in {^{\uu}SO(3,1)}
\eeq
Hence, the use of vierbeins reduces the gauge group $^{\uu}GL_{4}(\RR)$ of the theory to the group $^{\uu}SO(3,1)$. The latter possesses a spin cover, its elements will then be lifted to the spinor space \cite{SCH03}.

\newpage
\chapter{Fluctuation of the Dirac operator}

\section{Generalized Dirac operator}
In this chapter, we present the generalized Dirac operator, that we will simply call Dirac operator in the following.
However, in this chapter, we will make a distinction between the Dirac operator $\ds\ $ and the generalized Dirac operator $\Ds$. We will see that the word ``generalized'' has to be understood in the same way that General Relativity generalizes the Special Relativity theory.

\subsection{Short historical reminder}
Dirac's operator is born from the attempts to unify the Special Relativity theory and Quantum Mechanics. In the beginning of the latter, a dual mathematical formalisms did exist: Matrix Mechanics, initiated by W. Heisenberg and giving an operatorial viewpoint on Physics, and Wave Mechanics, conceptually closer to classical mechanics because deriving from an evolution equation similar to the fundamental principle of dynamics (the ordinary differential equation due to Newton $\overrightarrow{F} = m \overrightarrow{a}$), the Schrödinger equation:
\beq
\ \ \ \ \ i\hbar\ \frac{\pa}{\pa t}\ \psi_{t}(x^{i}) =\ -\ \frac{\hbar^{2}}{2m}\ {\nabla}^{2}\ \psi_{t}(x^{i}) \ \ \ \ \ \textrm{où}\ \nabla_{\!i} &:=& \frac{\pa}{\pa x^{i}} \\
\nnum  i &=& \{1,2,3\}
\eeq
\vspace{-0.7cm}

where $\psi_{t}(x^{i}) \in L^{2}\left(\RR^{3}\right)$ is the wavefunction, complex scalar function of four real variables. \\

\noi These two theories were in fact two different viewpoints of the same one that we call nowadays Quantum Mechanics. However, history made the choice to develop the one of Schrödinger and forgot, for some time, the Heisenberg viewpoint that Noncommutative Geometry brings back to light. \\
The attempts for unification between Quantum Mechanics and Special Relativity led physicists to find a version of the Schrödinger equation that is covariant under the Poincaré group. \\
Using the correspondence principle:
\beq
\bar{l}
E \longrightarrow\ i\hbar\ \pa_{t} \\
p_{i} \longrightarrow\ -\ i\hbar\ \pa_{x^{i}}
\ear\ \textrm{ou}\ \ P^{\mu}\ \longrightarrow\ i\hbar\ \pa^{\mu}
\eeq
and the relativistic invariant of the 4-vector energy-momentum:
\beq
P^{\mu} P_{\mu} = \frac{E^{2}}{c^{\,2}}\ -\ P^{2} = m^{2}c^{2}\ \ \ \longrightarrow\ \ -\ \hbar^{2}\
\pa^{\,\mu}\pa_{\mu}= -\ \hbar^{2}\ \Box
\eeq
lead to the formulation of the Klein-Gordon equation:
\beq
\left[\frac{\pa^{2}}{{c^{\,2\,}\pa t}^{2}}-\nabla^{2}+\frac{m^{2}c^{\,2}}{\hbar^{2}}\right]\ \psi(x^{\,\mu}) = \left[\Box+\frac{m^{2}c^{\,2}}{\hbar^{2}}\right]\ \psi(x^{\,\mu}) = 0
\eeq

where $\psi \in L^{2}\left(\RR^{4}\right)$ is still a complex scalar function.\\

\noi This equation only addresses spin 0 particles, and suffers from non-positive probability density. However, we can remark that this equation is covariant under the Poincaré group, it is in fact invariant ($\Box$, $m$ and $\psi$ being scalar quantities). \\
In order to obtain an equation able to describe half-integer spin particles, and to suppress the negative densities problem, Dirac tried to obtain an operator formally defined as the square root of the operator appearing in the Klein-Gordon equation. To this purpose, he realized the linearization of this equation (with $\hbar = c = 1$):
\beq
\left[\Box+m^{2}\right]\ \psi(x^{\,\mu}) = 0\ \ \ \Leftrightarrow\ \ \ \left[P^{\mu}P_{\mu}-m^{2}\right]\ \psi(x^{\nu}) =
0
\eeq
\beq
P^{\mu}P_{\mu}-m^{2} &=& \left(\alpha_{\mu}P^{\mu}+m\right)\ \left(\gamma_{\nu}P^{\nu}-m\right) \\
\nnum &=& \alpha_{\mu}P^{\mu}\ \gamma_{\nu}P^{\nu} - m^{2} + m P^{\mu} \left(\gamma_{\mu}-\alpha_{\mu}\right)
\eeq
where $\alpha_{\mu}$ and $\gamma_{\mu}$ are constants introduced to obtain scalar quantities, their indices being those of 4-vectors. Comparing this equation with the Klein-Gordon one, we observe that the absence of linear terms in $P^{\mu}$ imply: $\alpha_{\mu} = \gamma_{\mu}$; and we obtain the constraint on the $\gamma^{\mu}$ :
\beq
\gamma_{\mu}\gamma_{\nu}\ P^{\mu}P^{\nu} = P^{\mu}P_{\mu}
\eeq
Developing the components in this equation, and identifying the terms in the two sides of the equality lead to the constraint:
\beq
\left\{\gamma_{\mu},\gamma_{\nu}\right\} = 2\,\eta_{\mu\nu}
\eeq
We see that here appears the underlying Clifford algebra, this constraint implying also that the $\gamma^{\mu}$, have to be matrix elements. The simplest solution representing these $\gamma^{\mu}$ matrices, named Dirac matrices, is of dimension 4. \\
The Dirac equation reads:
\beq
\hspace{5cm} \left(\ds - m\right)\ \psi_{\sigma}(x^{\nu}) = 0\ \hspace{2cm} \textrm{où}\ \ \ds := i\gamma^{\mu} \pa_{\mu}
\eeq
and $\psi_{\sigma}\in L^{2}\left(\RR^{4}\right)$ is now a spinor and possesses four components. In this way is defined the Dirac operator $\ds$. \\
One observes that the Dirac equation is covariant under the Poincaré group, but the transformation law is different from the one of a scalar or a 4-vector, it corresponds to the one of a spinor. We observe that the Dirac operator itself is a scalar and thus invariant under the Poincaré group. \\
However, the theory formulated with this equation is not satisfactory, its main shortage being the violation of causality. Moreover, it should be necessary to obtain a theory allowing one to describe a changing number of particles in order to take into account for the disintegration phenomenon. A solid theory will be obtained later, with the Relativistic Quantum Field Theory.

\subsection{Dirac operator and Differential Geometry}
With the quest for unification, the General Relativity involved to generalize gauge theories, defined on Minkowskian spaces, to pseudo-Riemannian manifolds, generalization operated by Cartan (cf. Chapter 1). We will now be interested in incorporating the Dirac operator within the Differential Geometry formalism, which will be made in two steps:
\begin{itemize}
\item one incorporates the operator in a Minkowskian space (gauge theories)
\item one expresses the operator in a non-inertial coordinate system (equivalent to curved space)
\end{itemize}

\subsubsection{Incorporation within a Minkowskian space}
The Dirac operator $\ds := i\gamma^{\mu} \pa_{\mu}$ being a first order differential operator, it will be natural to define it from the exterior derivative $d$, whose action on a 0-form $\psi$ and a 1-form $\beta$ is :
\beq
d\psi_{\vert x} &=& \pa_{\mu}\psi(x)\ dx^{\mu} \\
\nnum d\beta_{\vert x} &=& \pa_{\mu}\beta_{\nu}(x)\ dx^{\mu}\wedge dx^{\nu}
\eeq
where $dx^{\mu}$ is an holonomic frame not necessarily inertial. \\
$\psi$ is now a 0-form belonging to $L^{2}\left(\CC^{4}\right)$ (spinor), located in a fiber whose structure group is $Spin(3,1)$. Th action of the covariant exterior derivative $D$, is given by the addition of a connection (1-form with values in the Lie algebra $spin(3,1)$) :
\beq
D\ :\ \Lambda^{0}M\ &\longrightarrow&\ T^{\ast}_{x}M\ \subset \Lambda^{1}M \\
\nnum \psi(x)\ &\longmapsto&\ \left(D\psi\right)_{\vert x} = D_{\vert x} \psi(x) = \left(\pa_{\mu}\psi(x)+\left(\rho\left(\lambda\right)\right)_{\mu}\psi(x)\right)\ dx^{\mu} {}\\
\nnum &&{}\hspace{1.4cm} = \left(\pa_{\mu}\psi(x)+ \frac{1}{4}\ \left(\omega(x)\right)_{ab\mu}\ \gamma^{a}\gamma^{b}\ \psi(x)\right)\ dx^{\mu}
\eeq
where $\lambda \in so(3,1)$ is an infinitesimal transformation and $\rho$ its representation (which corresponds to the lift $SO(3,1)\rightarrow Spin(3,1)$) on the spinor space. The $\gamma^{a}$ are the Dirac matrices and $\omega_{ab}$ the spin connection \cite{SCH02}.\\
One can express, by use of the vierbein ${\left(e(x)\right)^{a}}_{\mu}$ (cf. chapter 1), the Dirac operator in an orthonormal frame $e^{a}$ :
\beq
\left(D\psi\right)_{\vert x} = D_{\!\mu\vert x}\ \psi(x)\ dx^{\mu} = {\left(e^{-1}(x)\right)^{\mu}}_{a}\ D_{\!\mu\vert x}\ \psi(x)\ e^{a}
\eeq
We know thanks to the Gram-Schmidt theorem, that an orthonormal frame does exist on every chart of the manifold: a family of Dirac matrices  (where $\gamma^{a}$ are constant matrices and identical in every frame) is thus locally defined in the tangent space $T_{x}M$:
\beq
\gamma = \gamma^{a}\ e_{a\,\vert x}
\eeq
One can observe that this vector field, which is Dirac matrix valued, and the field of 1-forms $D\psi$ have inverse transformation laws under chart change and compensates each other.\\
We finally define the generalized Dirac operator, on a Minkowskian space and in an arbitrary coordinate system, by the operator:
\beq
\Ds\ :\ \Lambda^{0}M\ &\longrightarrow&\ \Lambda^{0}M \\
\nnum\psi\ &\longmapsto&\ \Ds\ \psi\ =\ i\left(\gamma\ , D\psi\right)\ =\ i\gamma^{a}\ {\left(e^{-1}(x)\right)^{\mu}}_{b}\ D_{\!\mu\vert x}\ \psi(x)\ \left(e_{a}\ , e^{b}\right) {}\\
\nnum {}&&\hspace{1.2cm} =\ i\gamma^{a}\ {\left(e^{-1}(x)\right)^{\mu}}_{a}\ \left(\pa_{\mu}+ \frac{1}{4}\ \left(\omega(x)\right)_{ab\mu}\ \gamma^{a}\gamma^{b}\right) \psi
\eeq
where $\left(\cdot\ ,\cdot\right)$ is the duality product.\\
It is also shown that imposing a covariance under the diffeomorphisms of the manifold lead to the appearance of the vierbein and the spin connection (\cite{FRI01}).

\section{Conjecture on the fluctuation}
{\bfseries Theorem ({\small Moser})} :\ \ \ Be $M$ a differentiable manifold of dimension $n$ without boundary,
orientable, connected and compact. Be $\omega$, $\eta \in \Omega^{n}$ two volume forms with
\beq
\int_{M} \omega = \int_{M} \eta
\eeq
There exists a diffeomorphism $\varphi$: $M \longrightarrow M$ such that $\varphi^{\ast} \omega = \eta$. \\

{\bfseries Conjecture} :\ \ \ On a spin Riemannian manifold, every Dirac operator $\Df$ can be obtained from an arbitrary Dirac operator $\Ds$, as a finite linear combination of coordinate transformations applied to $\Ds$.
\noi The Dirac operators $\Df$ are called fluctuated Dirac operators. \\

We will show here after, far from giving the proof of this conjecture, that a fluctuation method leads to express a fluctuated Dirac operator $\Df$, in a frame derived from a coordinate system $x$, by the general expression:
\beq
\Df_{\vert x} =\ \frac{\sum\limits_{i\, =\, 1}^{s}\
\alpha_{i}\ {^{\varphi_{i}}\!\Ds}_{\vert \varphi_{i}(x)}} {\Big{\vert}\det
\left[\ {\left(e(x)\right)^{a}}_{\mu} \, \ \sum\limits_{i\, =\, 1}^{s}\
\alpha_{i}\ {\left(^{\varphi_{i}}e^{-1}(\varphi_{i}(x))\right)^{\nu}}_{c}\
{\left(J^{-1}_{\varphi_{i}}(x)\right)^{\mu}}_{\nu}\right]\Big{\vert}^{\frac{1}{n-1}}}
\eeq
\beq
\nnum \textrm{where} &&\Df_{\vert x}\ \textrm{is written in the coordinate system in which the fluctuation is operated.} \\
\nnum &&\alpha_{i}\in\RR\ \ \textrm{are constants.} \\
\nnum &&\left\{\varphi_{i}\right\}_{i\, =\, 1}^{s}\ \textrm{is a finite set of diffeomorphisms in the connected component}\\
\nnum &&\ \ \ \ \ \ \textrm{of identity and such that:}\ \ x_{\varphi} = \varphi(x)\ . \\
\nnum &&{^{\varphi_{i}}\!\Ds} = L\left(\varphi_{i}\right) \Ds\ L\left(\varphi_{i}\right)^{-1}\ \ \textrm{where $\Ds$ is the initial Dirac operator.} \\
\nnum &&{\left(J_{\varphi_{i}}(x)\right)^{\mu}}_{\nu}\ \ \textrm{is the jacobian of the diffeomorphism $\varphi_{i}$.} \\
\nnum &&{(e(x))^{a}}_{\mu}\ \ \textrm{is the vierbein of the initial Dirac operator.}
\eeq

In the same way as the General Relativity gives to the metric a role of dynamical variable, solution of an evolution equation (the Einstein equation), Noncommutative Geometry does the same by giving to the Dirac operator an equivalent role to that of the metric. However, contrary to the metric case, linear combinations of Dirac operators are permitted. \\

\noi The dynamical configuration space $\ff$ of every Dirac operators is defined by
\beq
\ff := \left\{\ \Df\ \ \Big{\vert}\ \ \Df_{\vert x} =\ \frac{\sum\limits_{i\, =\, 1}^{s}\
\alpha_{i}\ {^{\varphi_{i}}\!\Ds}_{\vert \varphi_{i}(x)}} {\Big{\vert}\det
\left[\ {\left(e(x)\right)^{a}}_{\mu} \, \ \sum\limits_{i\, =\, 1}^{s}\
\alpha_{i}\ {\left(^{\varphi_{i}}e^{-1}(\varphi_{i}(x))\right)^{\nu}}_{c}\
{\left(J^{-1}_{\varphi_{i}}(x)\right)^{\mu}}_{\nu}\right]\Big{\vert}^{\frac{1}{n-1}}} \right\}
\eeq

This dynamical configuration space is, so far, restricted to Dirac operators describing spaces without torsion. The results obtained in this report (i.e. generation of torsion from initially flat spaces) shows the necessity to extend this configuration space to Dirac operators describing spaces with torsion also. This requires to be able to determine the spin connection of such an operator.

\newpage
\section{Fluctuated Dirac operator}
In this section, we give the general form of the fluctuated Dirac operator $\Df$, expressed in terms of the fluctuated vierbein and the fluctuated spin connection. In the first part, the fluctuation is operated by using a single operator sum, and will then be carried out in the general case of a fluctuated Dirac operator by means of multiple sums. \\
The process which led to this fluctuation method has implied to define a sum of operators deduced from the scalar product defining the Dirac action. \\

In the following, we use an euclidian metric in order to allow the computation of its square root. This is a necessary condition if one wants to work in the symmetric gauge, for which the vierbein is defined as the square root of the metric.

\subsection{Fluctuation : a preliminary description}
\noi We consider the frame of a particular observer $O$ endowed with a coordinate system $x^{\mu}$. In this frame, the Dirac action of an arbitrary field $\psi$ reads:
\beq
\label{Scal01}\left( \psi,\Ds \psi \right) &=& \int_{\RR^{n}}{\dnx {\vert \det e\,\vert}_{\vert x}\ \psi^{*}(x) \Ds_{\vert x}\ \psi(x)} \\
\nnum &=& \int_{\RR^{n}}{\dnx {\vert \det e\,\vert}_{\vert x}\ \psi^{*}(x) \ i{\left(e^{-1}(x)\right)^{\mu}}_{c}\ \gamma^{c} \left(\frac{\pa}{\pa x^{\mu}}+\frac{1}{4} \left(\omega(x)\right)_{ab\mu}\gamma^{a}\gamma^{b}\right) \psi(x)}
\eeq
Let as draw attention to an expression that will appear more fundamental than the Dirac operator itself for the fluctuation purpose:
\beq
{\vert \det e\,\vert}_{\vert x}\ \Ds_{\vert x} = {\vert \det e\,\vert}_{\vert x}\
i{\left(e^{-1}(x)\right)^{\mu}}_{c}\ \gamma^{c} \left(\frac{\pa}{\pa x^{\mu}}+\frac{1}{4}
\left(\omega(x)\right)_{ab\mu}\gamma^{a}\gamma^{b}\right)
\eeq

\noi In the following, we consider the frame of an observer $O_{\varphi}$ endowed with a coordinate system $x_{\svphi}$, deduced from the previous one by a diffeomorphism $\varphi$ in the connected component of identity (for which $\det J_{\svphi} > 0$) such that:
\vspace{-0.5cm}
\beq
\nnum (x_{\svphi})^{\mu} &=& \varphi^{\mu}\!\left( x \right) \\
\textrm{and thus}\ \ \ x^{\mu} &=& \left(\varphi^{-1}(x_{\svphi})\right)^{\mu}\\
\nnum \textrm{with} \ \ \ (dx_{\svphi})^{\mu} &=& \frac{\pa \varphi^{\mu}\!\left( x \right)}{\pa x^{\nu}}\
dx^{\nu} =: {(J_{\svphi}(x))^{\mu}}_{\nu}\ dx^{\nu}
\eeq
In this frame, the Dirac operator $^{\varphi\!\!}\Ds$, deduced from the previous one by the lift of the diffeomorphism
\vspace{-0.5cm}
\beq
^{\varphi\!\!}\Ds = L(\varphi)\Ds L(\varphi)^{-1}
\eeq
allows to define the action of the field $\psi_{\svphi} = L\left(\varphi\right)\psi$:
\beq
\label{Fluc001}\sp\left( \psi_{\svphi},^{\varphi\!\!\!}\Ds \psi_{\svphi} \right) &=& \int_{\RR^{n}}{\sp d^{n}\! x_{\svphi} {\vert \det
{^{\varphi\!\!}e\,}\vert}_{\vert x_{\svphi}} \ \psi_{\svphi}^{*}(x_{\svphi}) }\\
\nnum &&\ \ \ \ \ \ \ \ i{\left(^{\varphi\!\!}e^{-1}(x_{\svphi})\right)^{\mu}}_{c}\gamma^{c} \left(\frac{\pa}{\pa {x_{\svphi}}^{\mu}}+\frac{1}{4} \left(^{\varphi\!\!}\omega(x_{\svphi})\right)_{ab\mu}\gamma^{a}\gamma^{b}\right) \psi_{\svphi}(x_{\svphi})
\eeq
for which we know that $\left( \psi_{\svphi},^{\varphi\!\!\!}\Ds \psi_{\svphi} \right) = \left( \psi,\Ds\psi\right)$.
\newpage
\noi The  theorem of change of variables allows to write:
\beq
\label{Fluc002}(x_{\svphi})^{\mu} &\longrightarrow& (x_{\svphi}(x))^{\mu} = \varphi^{\mu}\!\left( x \right) \\
\nnum {dx_{\svphi}}^{\mu} &\longrightarrow& d \left[ (x_{\svphi}(x))^{\mu} \right] = {(J_{\svphi}(x))^{\mu}}_{\nu}\
dx^{\nu} \\
\nnum \Longrightarrow d^{n}\! x_{\svphi} &\longrightarrow& d^{n}\! (x_{\svphi}(x)) = \det(J_{\svphi})_{\vert x}\ d^{n}\! x
\eeq
where $J_{\svphi}$ is the jacobian of the diffeomorphism $\varphi$. We can thus express the action in the coordinates $x^{\mu}$ of the previous frame. \\
At this point, we have to put the emphasis on the difference between this change of variables, which expresses a single term in two different coordinate systems linked by a diffeomorphism $\varphi$, and the diffeomorphism lift, which generates a rotation in the Hilbert space of spinors and engenders a real difference between the two Dirac operators $\Ds$ et $^{\varphi\!\!}\Ds$. With our notations:
\beq
^{\varphi\!\!}\Ds_{\vert x_{\svphi}} &\sim& {^{\varphi\!\!}\Ds}_{\vert\varphi(x)} \\
\nnum &\nsim& \Ds_{\vert\varphi^{-1}\left(x_{\svphi}\right)}
\eeq
This notation can be resumed by: $\ \ \ \ ^{\textrm{lift}\!}\Ds_{\vert\textrm{coordonnées}}$ .\\

\noi With this remark, we can write equation (\ref{Fluc001}), by means of the change of variables (\ref{Fluc002}), in the initial coordinates $x^{\mu}$ :
\beq
\label{Scal02}\left( \psi_{\svphi},^{\varphi\!\!\!}\Ds \psi_{\svphi} \right) &=& \int_{\RR^{n}}{\sp \det(J_{\svphi})_{\vert x}\ d^{n}\! x  {\vert \det
{^{\varphi\!\!}e\,}\vert}_{\vert
\varphi(x)}\ \psi_{\svphi}^{*}(\varphi(x))} \\
\nnum &&\ \ \ \ \ \ \ \ i{\left(^{\varphi\!\!}e^{-1}(\varphi(x))\right)^{\mu}}_{c}\ \gamma^{c} \left({(J^{-1}_{\svphi}(x))^{\nu}}_{\mu}\
\frac{\pa}{\pa {x}^{\nu}}+\frac{1}{4}
\left(^{\varphi\!\!}\omega(\varphi(x))\right)_{ab\mu}\gamma^{a}\gamma^{b}\right)
\psi_{\svphi}(\varphi(x))
\eeq

\noi Here again, we want to highlight the term:
\beq
\label{Fluc003}&&\sp\sp\sp \det(J_{\svphi})_{\vert x}\ {\vert \det {^{\varphi\!\!}e\,}\vert}_{\vert
\varphi(x)}\
^{\varphi\!\!}\Ds_{\vert\varphi(x)} = \\
&&\nnum \det(J_{\svphi})_{\vert x}\ {\vert \det {^{\varphi\!\!}e\,}\vert}_{\vert \varphi(x)}\
i{\left(^{\varphi\!\!}e^{-1}(\varphi(x))\right)^{\mu}}_{c}\ \gamma^{c} \left({(J^{-1}_{\svphi}(x))^{\nu}}_{\mu}\
\frac{\pa}{\pa {x}^{\nu}}+\frac{1}{4} \left(^{\varphi\!\!}\omega(\varphi(x))\right)_{ab\mu}
\gamma^{a}\gamma^{b}\right)
\eeq

We have chosen to express the vierbeins ${e^{a}}_{\mu}$ and ${\left(^{\varphi\!}e\right)^{a}}_{\mu}$ in the symmetric gauge and, in that case, we have written in an equivalent way the volume form:
\beq
dV_{n} = \sqrt{\vert\det g_{..}\vert}_{\vert x}\ \ \ \ \dnx = \vert\det e\,\vert_{\vert x}\ \ \ \ \dnx
\eeq
Moreover, we know by the same reasoning as in (\ref{Conj004}), that two metric tensors $g_{\mu\nu}$ and ${^{\varphi\!}g_{\mu\nu}}$ are related by:
\beq
\left(^{\varphi\!\!}e\left(\varphi(x)\right)\right)_{\mu\nu} = \sqrt{\left({^{\varphi\!}g}\left(\varphi(x)\right)\right)}_{\mu\nu} = \sqrt{\left(J^{-T}_{\svphi}(x)\ g(x)\ J^{-1}_{\svphi}(x)\right)}_{\mu\nu}
\eeq
and the term $\sqrt{\vert\det g_{..}\,\vert}$, or here the term $\vert\det e\,\vert$, transforms with the jacobian under a change of variables:
\beq
\det(J_{\svphi})_{\vert x}\ {\vert \det {^{\varphi\!\!}e\,}\vert}_{\vert
\varphi(x)}\ {^{\varphi\!\!}\Ds}_{\vert\varphi(x)} &=& \det(J_{\svphi})_{\vert x}\ \left\vert\det\left(\sqrt{\left(J^{-T}_{\svphi}\ g\ J^{-1}_{\svphi}\right)}_{..}\right)\right\vert_{\vert x}\ {^{\varphi\!\!}\Ds}_{\vert\varphi(x)} \\
\nnum &=& \det(J_{\svphi})_{\vert x}\ \left(\det \sqrt{g}_{..}\right)_{\vert x}\ \det(J^{-1}_{\svphi})_{\vert x}\ {^{\varphi\!\!}\Ds}_{\vert\varphi(x)} \\
\nnum &=& \vert\det e\,\vert_{\vert x}\ {^{\varphi\!\!}\Ds}_{\vert\varphi(x)}
\eeq
Hence, the only term keeping trace of the diffeomorphism in equation (\ref{Fluc003}) is the Dirac operator. \\

From the scalar products appearing in the actions (\ref{Scal01}) and (\ref{Scal02}), we will determine the fluctuated Dirac operator $\Df$ in the frame related to the observer $O$. To this purpose, we define $\Df$ as the operator determining the action of the field $\psi$ in the frame $O$ after fluctuation.
The invariant measure from the action integral is then expressed with the fluctuated vierbein ${(\ef\ )^{a}}_{\mu}$.
\beq
\left( \psi,\Df \psi \right) &=& \int_{\RR^{n}}{\dnx {\vert \det \ef\,\vert}_{\vert x}\ \psi^{*}(x)\ \Df_{\vert x}\ \psi(x)} \\
\nnum &=& \int_{\RR^{n}}{\dnx {\vert \det \ef\,\vert}_{\vert x}\ \psi^{*}(x) \ i{\left(\ef^{-1}(x)\right)^{\mu}}_{c}\gamma^{c} \left(\frac{\pa}{\pa x^{\mu}}+\frac{1}{4} \left(\omegaf(x)\right)_{ab\mu}\gamma^{a}\gamma^{b}\right) \psi(x)}
\eeq

\noi In order to ensure that the resulting Dirac operator be hermitian, we have to sum, not the Dirac operators themselves, but the entire terms found under the integral :\\
\beq
{\vert \det \ef\,\vert}_{\vert x}\ \Df_{\vert x} &=& {\vert \det e\,\vert}_{\vert x}\ \Ds_{\vert x} +
\det(J_{\svphi})_{\vert x}\ {\vert \det {^{\varphi\!\!}e\,}\vert}_{\vert \varphi(x)}\ ^{\varphi\!\!}\Ds_{\vert
\varphi(x)}\\
\nnum &=& {\vert \det e\,\vert}_{\vert x}\ \left(\Ds_{\vert x} + {^{\varphi\!\!}\Ds}_{\vert
\varphi(x)} \right)
\eeq
\\
\noi We can now determine the fluctuated vierbein ${(\ef(x))^{a}}_{\mu}$ and the fluctuated spin connection $(\omegaf(x))_{ab\mu}$ belonging to the frame $O$.
After having explicitly expressed the Dirac operators, we can identify the following terms:
\beq
\label{fluct01}{\vert \det \ef\,\vert}_{\vert x}\ {\left(\ef^{-1}(x)\right)^{\mu}}_{c} &=& {\left(E^{-1}\right)^{\mu}}_{c}\\
\nnum &=& {\vert \det e\,\vert}_{\vert x}\ \left[{\left(e^{-1}(x)\right)^{\mu}}_{c} +  {\left(^{\varphi\!\!}e^{-1}(\varphi(x))\right)^{\nu}}_{c}
{\left(J^{-1}_{\svphi}\right)^{\mu}}_{\nu}\right] \\
\nnum \\
\label{fluct02}{\vert \det \ef\,\vert}_{\vert x}\ {\left(\ef^{-1}(x)\right)^{\mu}}_{c}
\left(\omegaf(x)\right)_{ab\mu} &=& \left(\Omega_{\scriptscriptstyle{E}}\right)_{abc} \\
\nnum &=& {\vert \det e\,\vert}_{\vert x}\ \left[{\left(e^{-1}(x)\right)^{\mu}}_{c} \left(\omega\left(x\right)\right)_{ab\mu}\right. {} \\
\nnum &&{}\ \ \ \ \ \ \ \ \ \ \ \ \ \left. +\ {\left(^{\varphi\!\!}e^{-1}(\varphi(x))\right)^{\mu}}_{c}
\left(^{\varphi\!\!}\omega(\varphi(x))\right)_{ab\mu}\right]
\eeq

\noi We can deduce the expression of the term ${\vert \det \ef\,\vert}_{\vert x}$ by computing the determinant of equation (\ref{fluct01}).
\vspace{-0.5cm}
\beq
&&\sp\sp \det [(\ref{fluct01})] \longrightarrow \left(\det \ef\,\right)^{n-1}_{\vert x}
\eeq
\vspace{-0.6cm}
\beq
\nnum &&\sp\sp\Rightarrow {\vert \det \ef\,\vert}_{\vert x} = \vert \det \left( E^{-1}(x) \right)\vert^{\frac{1}{n-1}} =
\vert \det \left(E(x) \right)\vert^{\frac{-1}{n-1}}
\eeq

\noi We can thus obtain the expressions of the fluctuated vierbein and the fluctuated spin connection:
\beq
{\left(\ef (x) \right)^{a}}_{\mu} &=& \vert \det \left(E (x) \right)\vert^{\frac{-1}{n-1}}\ {\left(E(x)\right)^{a}}_{\mu} \\
\left(\omegaf(x)\right)_{ab\mu}\!\!\! &=& \left(\Omega_{\scriptscriptstyle{E}}(x)\right)_{abc}\
{\left(E(x)\right)^{c}}_{\mu}
\eeq

\indent The fluctuated vierbein and spin connection are now written in an abstract manner, and are functions of terms
${\left(E(x)\right)^{a}}_{\mu}$ and $\left(\Omega_{\scriptscriptstyle{E}}(x)\right)_{abc}$. The explicit form of these terms allows one to easily generalize the fluctuation method to an arbitrary sum of diffeomorphisms. \\

\subsection{Fluctuation : general case}
We consider a finite set of diffeomorphisms $\left \{\varphi_{i}\right\}_{\scriptscriptstyle i\, =\,
1}^{\scriptscriptstyle s}$. We restrict this set to diffeomorphisms that preserve orientation of frames in the tangent fiber of the bundle: $\det \left( J_{\varphi_{i}} \right)>0$. These diffeomorphisms are thus in the component connected to identity.
However, we stress that unicity of lifts is only ensured for diffeomorphisms infinitesimally close to the identity \cite{IOC01}. \\
\noi Each of these diffeomorphisms applies the coordinates $x^{\mu}$ from an observer $O$ to the coordinates of another observer $O_{\varphi_{i}}$ :
\beq
\nnum (x_{\varphi_{i}})^{\mu} &=& \varphi_{i}^{\mu}\!\left( x \right) \\
x^{\mu} &=& \left(\varphi_{i}^{-1}(x_{\varphi_{i}})\right)^{\mu}\\
\nnum (dx_{\varphi_{i}})^{\mu} &=& \frac{\pa \varphi_{i}^{\mu}\!\left( x \right)}{\pa x^{\nu}}\
dx^{\nu} =: {(J_{\varphi_{i}}(x))^{\mu}}_{\nu}\ dx^{\nu}
\eeq
To each observer $O_{\varphi_{i}}$ corresponds a Dirac operator $^{\varphi_{i}}\!\Ds~=~L\left(\varphi_{i}\right) \Ds\ L\left(\varphi_{i}\right)^{-1}$, written with coordinates $x_{\varphi_{i}}$, and an associated Dirac action defined by the appropriate scalar product :
\beq
\left( \psi_{\varphi_{i}},^{\varphi_{i}}\!\!\Ds \psi_{\varphi_{i}} \right) =  \int_{\RR^{n}}{\dnx_{\varphi_{i}} {\vert \det {^{\varphi_{i}}\!e}\,\vert}_{\vert x_{\varphi_{i}}}\ \psi_{\varphi_{i}}^{*}(x_{\varphi_{i}})\ ^{\varphi_{i}}\!\Ds_{\vert x_{\varphi_{i}}}\ \psi_{\varphi_{i}}(x_{\varphi_{i}})}
\eeq
\vspace{-0.76cm}
\beq
\nnum = \int_{\RR^{n}}{\dnx_{\varphi_{i}} {\vert \det {^{\varphi_{i}}\!e}\,\vert}_{\vert x_{\varphi_{i}}}\
\psi_{\varphi_{i}}^{*}(x_{\varphi_{i}}) \
i{\left({^{\varphi_{i}}e}^{-1}(x_{\varphi_{i}})\right)^{\mu}}_{c}\gamma^{c} \left(\frac{\pa}{\pa
x_{\varphi_{i}}^{\mu}}+\frac{1}{4}
\left({^{\varphi_{i}}\omega}(x_{\varphi_{i}})\right)_{ab\mu}\gamma^{a}\gamma^{b}\right)
\psi_{\varphi_{i}}(x_{\varphi_{i}})}
\eeq
The theorem of change of variables  allows to write the operator $^{\varphi_{i}}\!\Ds$ in the coordinates $x^{\mu}$, and the corresponding jacobian
changes the differential part of the operator:
\beq
\left( \psi_{\varphi_{i}},^{\varphi_{i}}\!\!\Ds \psi_{\varphi_{i}} \right) &=& \int_{\RR^{n}}{\dnx
\det(J_{\svphi_{i}})_{\vert x}\ {\vert \det {^{\varphi_{i}}\!e}\,\vert}_{\vert \varphi_{i}(x)}\
\psi_{\varphi_{i}}^{*}(\varphi_{i}(x))\ ^{\varphi_{i}}\!\Ds_{\vert \varphi_{i}(x)}\
\psi_{\varphi_{i}}(\varphi_{i}(x))} \\
\nnum &=& \int_{\RR^{n}}{\dnx
\vert \det e\,\vert_{\vert x}\
\psi_{\varphi_{i}}^{*}(\varphi_{i}(x))\ ^{\varphi_{i}}\!\Ds_{\vert \varphi_{i}(x)}\
\psi_{\varphi_{i}}(\varphi_{i}(x))}
\eeq
The linear combination with coefficients $\alpha_{i} \in \RR$ of terms appearing behind the integral sign (now unique with integration variable $x$, belonging to the frame of observer $O$) leads to:
\beq
\label{RelDFluct}{\vert \det \ef\,\vert}_{\vert x}\ \Df_{\vert x} = \vert \det e\, \vert_{\vert x}\ \sum_{i\, =\, 1}^{s}\
\alpha_{i}\ {^{\varphi_{i}}\!\Ds}_{\vert \varphi_{i}(x)}
\eeq
\noi We can express the terms ${\left(E(x)\right)^{a}}_{\mu}$ and
 $\left(\Omega_{\scriptscriptstyle{E}}(x)\right)_{abc}$ in the general case :
\beq
\label{Fluc006}{\left(E^{-1}(x) \right)^{\mu}}_{a} &=&\ \vert \det e\, \vert_{\vert x}\ \sum_{i\, =\, 1}^{s}\
\alpha_{i}\ {\left(^{\varphi_{i}\!}e^{-1}(\varphi_{i}(x))\right)^{\nu}}_{a}\
{\left(J^{-1}_{\varphi_{i}}(x)\right)^{\mu}}_{\nu} \\
\left(\Omega_{\scriptscriptstyle{E}}(x)\right)_{abc}\ &=&\ \vert \det e\, \vert_{\vert x}\ \sum_{i\, =\, 1}^{s}\ \alpha_{i}\
{\left(^{\varphi_{i}\!}e^{-1}(\varphi_{i}(x))\right)^{\mu}}_{c}\
\left(^{\varphi_{i}\!}\omega(\varphi_{i}(x))\right)_{ab\mu}
\eeq

\noi with which we express the fluctuated spin connection, vierbein and the absolute value of its determinant:
\beq
\left(\omegaf(x)\right)_{ab\mu}\!\!\! &=& \left(\Omega_{\scriptscriptstyle{E}}(x)\right)_{abc}\
{\left(E(x)\right)^{c}}_{\mu} \\
\label{Fluc005}{\left(\ef (x) \right)^{a}}_{\mu} &=& \vert \det \left(E (x) \right)\vert^{\frac{-1}{n-1}}\ {\left(E(x)\right)^{a}}_{\mu} \\
{\vert \det \ef\,\vert}_{\vert x}\! &=& \vert \det \left(E(x) \right)\vert^{\frac{-1}{n-1}}
\eeq
\\
\noi where the relatively simple expression of the determinant allows one to write:
\beq
\vert \det \ef\,\vert_{\vert x}\!\! &=& \vert \det e\, \vert_{\vert x}^{\frac{n}{n-1}}\ \ \left\vert\det
\left[\ \sum_{i\, =\, 1}^{s}\
\alpha_{i}\ {\left(^{\varphi_{i}\!}e^{-1}(\varphi_{i}(x))\right)^{\nu}}_{c}\
{\left(J^{-1}_{\varphi_{i}}(x)\right)^{\mu}}_{\nu} \right]\right\vert^{\frac{1}{n-1}}
\eeq

\noi From (\ref{RelDFluct}), this lead to express the fluctuated Dirac operator as follows:
\beq
\Df_{\vert x} =\ \frac{\sum\limits_{i\, =\, 1}^{s}\
\alpha_{i}\ {^{\varphi_{i}}\!\Ds}_{\vert \varphi_{i}(x)}} {\Big{\vert}\det
\left[\ {\left(e(x)\right)^{a}}_{\mu} \, \ \sum\limits_{i\, =\, 1}^{s}\
\alpha_{i}\ {\left(^{\varphi_{i}}e^{-1}(\varphi_{i}(x))\right)^{\nu}}_{c}\
{\left(J^{-1}_{\varphi_{i}}(x)\right)^{\mu}}_{\nu}
\right]\Big{\vert}^{\frac{1}{n-1}}}
\eeq

\beq
\nnum \textrm{where} &&\Df_{\vert x}\ \textrm{is expressed with the coordinates of the frame in which is operated the fluctuation.} \\
\nnum &&\alpha_{i}\in\RR\ \ \textrm{are constants.} \\
\nnum &&\left\{\varphi_{i}\right\}_{i\, =\, 1}^{s}\ \textrm{is a finite set of diffeomorphisms infinitesimally close to}\\
\nnum &&\ \ \ \ \ \ \textrm{identity and such that :}\ \ x_{\varphi} = \varphi(x)\ . \\
\nnum &&{^{\varphi_{i}}\!\Ds} = L\left(\varphi_{i}\right) \Ds\ L\left(\varphi_{i}\right)^{-1}\ \ \textrm{where $\Ds$ is the original Dirac operator.} \\
\nnum &&{\left(J_{\varphi_{i}}(x)\right)^{\mu}}_{\nu}\ \ \textrm{is the jacobian of the diffeomorphism $\varphi_{i}$.} \\
\nnum &&{(e(x))^{a}}_{\mu}\ \ \textrm{is the vierbein of the original Dirac operator.}
\eeq
Hence, we have obtained the general form of the fluctuated Dirac operator, expressed in terms of diffeomorphisms $\varphi_{i}$ (through their jacobian $J_{\varphi_{i}}(x)$), as well as the fluctuated vierbeins and spin connections (resulting from the lifts of diffeomorphisms $L\left(\varphi_{i}\right)$ applied to $\Ds$).
\beq
\Df = \Df\left(e,\omega,\alpha_{i},\varphi_{i},^{\varphi_{i}}\!e,^{\varphi_{i}}\!\omega\right)
\eeq
It is worth noting that explicit calculation of an arbitrary fluctuated Dirac operator requires to determine the spin connection $\left(^{\varphi_{i}\!}\omega(\varphi_{i}(x))\right)_{ab\mu}$, also in the case where the original space (i.e. before applying the lift $L\left(\varphi_{i}\right)$) has torsion.

\newpage
\chapter[Examples of fluctuation]{Examples of fluctuation}
In this chapter, after a preliminary discussion concerning the constraint on the torsion of space induced by the hermiticity of the Dirac operator, we develop two examples of fluctuation. We use coordinate transformations between two frames in flat spaces. First, a 2-dimensional example using the cartesian/polar coordinate transformation, then the same in three dimensions.

\section{Hermiticity of Dirac operator and torsion of space}
In the following, we show that hermiticity of the Dirac operator (required to ensure real-valued energy eigenstates) implies a constraint on the torsion of space-time. This property will allow to test the fluctuation method by making sure that the resulting operator fulfill this requirement. By simplicity we restrict this derivation to a 2-dimensional case, which corresponds to the first example. However, this derivation can be generalized to an arbitrary $n$-dimensional case. In the case we study here, the following relations hold:
\beq
^{*} &\Longleftrightarrow& ^{\dagger} \\
g_{\mu\nu} &=& \delta_{\mu\nu} \ =\ 1_{2} \\
\gamma^{\mu} &=& \sigma^{\mu} \ ,\  \textrm{implying hermiticity of }\gamma \textrm{ matrices} \\
(\gamma^{\mu})^{\dagger} &=& \gamma^{\mu} \\
\lbrace \gamma^{a},\gamma^{b} \rbrace &=& 2 g^{ab} \ = \ 2 \delta^{ab}
\eeq
where the matrix $\gamma^{3} = \sigma^{3}$ is the chirality operator.\\

\noi Writing the matrix element:
\beq
\nnum \left( \phi,\Ds\psi\right)&=&\int_{\RR^{2}}{d^{2}x \sqrt{\vert det\ g_{(x)..}\vert}\ \phi^{\dagger}_{(x)}\ \Ds_{\vert x}\  \psi_{(x)}} \\
&=&\int{d^{2}x \sqrt{\vert det\ g_{..}\vert} \ \phi^{\dagger} \ i(e^{-1})^{\mu}_{\,\ c}\gamma^{c} \left(\frac{\partial}{\partial x^{\mu}}+\frac{1}{4}\omega_{ab\mu}\gamma^{a}\gamma^{b}\right) \psi} \\
\nnum &=&\int{d^{2}x \vert\det e\,\vert_{\vert x} \ \phi^{*} \ i(e^{-1})^{\mu}_{\,\ c}\gamma^{c} \frac{\partial\psi}{\partial x^{\mu}}}\ +\ \int{d^{2}x \vert\det e\,\vert_{\vert x} \  \phi^{*} \ i(e^{-1})^{\mu}_{\,\ c}\gamma^{c}\frac{1}{4}\omega_{ab\mu}\gamma^{a}\gamma^{b}\psi} \\
\nnum &=&(I)+(II)
\eeq

\noi The spin connection $\omega(e, \partial e)$ satisfy the relations:
\beq
\omega_{ab}\in so(2) \Longrightarrow  (\omega^{\dagger})_{ba} = (\omega^{T})_{ba} = -\omega_{ab}
\eeq
\noindent The inverse vierbein $(e^{-1})^{\mu}_{\,\ c}$ being a real symmetric matrix :
\beq
((e^{-1})^{\mu}_{\,\ c})^{\dagger}=((e^{-1})^{\mu}_{\,\ c})^{T}=(e^{-1})^{\mu}_{\,\ c}
\eeq
\\

\noindent $\bullet$ Detailed calculation of the first part:
\beq
(I) = \int{d^{2}x \vert\det e\,\vert_{\vert x} \ \phi^{\dagger} \ i(e^{-1})^{\mu}_{\,\ c}\gamma^{c}
\frac{\partial\psi}{\partial x^{\mu}}}
\eeq
\noindent An integration by parts gives:
\beq
\nonumber \frac{\partial}{\partial x^{\mu}}\left[ \vert\det e\,\vert_{\vert x} \phi^{\dagger} (e^{-1})^{\mu}_{\,\ c}\gamma^{c} \psi \right] &=& \vert\det e\,\vert_{\vert x} \frac{\partial \phi^{\dagger}}{\partial x^{\mu}} (e^{-1})^{\mu}_{\,\ c}\gamma^{c} \psi + \vert\det e\,\vert_{\vert x} \phi^{\dagger} (e^{-1})^{\mu}_{\,\ c}\gamma^{c} \frac{\partial\psi}{\partial x^{\mu}} \\
&\ & +\frac{\partial}{\partial x^{\mu}}\left[ \vert\det e\,\vert_{\vert x} (e^{-1})^{\mu}_{\,\ c} \right]
\phi^{\dagger}\gamma^{c}\psi
\eeq

\beq
\nonumber (I)&=& i\int{d^{2}x \ \frac{\partial}{\partial x^{\mu}}\left[ \vert\det e\,\vert_{\vert x} \phi^{\dagger} (e^{-1})^{\mu}_{\,\ c}\gamma^{c} \psi \right]} - i\int{d^{2}x \vert\det e\,\vert_{\vert x} \ \frac{\partial \phi^{\dagger}}{\partial x^{\mu}} (e^{-1})^{\mu}_{\,\ c}\gamma^{c} \psi}\\
&\ & -i\int{d^{2}x \ \frac{\partial}{\partial x^{\mu}}\left[ \vert\det e\,\vert_{\vert x} (e^{-1})^{\mu}_{\,\ c} \right] \phi^{\dagger}\gamma^{c}\psi}\\
\nonumber &=& \int{d^{2}x \vert\det e\,\vert_{\vert x} \ (i)^{\dagger} (e^{-1})^{\mu}_{\,\ c} \frac{\partial \phi^{\dagger}}{\partial x^{\mu}} (\gamma^{c})^{\dagger} \ \psi} - i\int{d^{2}x \frac{\partial}{\partial x^{\mu}}\left[ \vert\det e\,\vert_{\vert x} (e^{-1})^{\mu}_{\,\ c} \right] \phi^{\dagger}\gamma^{c}\psi}
\eeq
\noindent where the simplification of the boundary term in the first equality is obtained for sufficiently fast decreasing fields ensuring the fields are null at infinity.\\

\noindent Hence, the term $(I)$ becomes:
\beq
(I)&=& \int{d^{2}x \vert\det e\,\vert_{\vert x} \left(i (e^{-1})^{\mu}_{\,\ c} \gamma^{c} \frac{\partial \phi}{\partial x^{\mu}} \right)^{\dagger} \psi} \\
\nonumber &\ & -i\int{d^{2}x \frac{\partial}{\partial x^{\mu}}\left[ \vert\det e\,\vert_{\vert x} (e^{-1})^{\mu}_{\,\ c} \right] \phi^{\dagger}\gamma^{c}\psi}
\eeq
\\

\noi $\bullet$ Detailed calculation of the second part:
\beq
\nnum (II) &=& \int{d^{2}x \vert\det e\,\vert_{\vert x} \ \phi^{\dagger} \ i(e^{-1})^{\mu}_{\,\ c}\gamma^{c}\frac{1}{4}\omega_{ab\mu}\gamma^{a}\gamma^{b} \ \psi} \\
&=& \int{d^{2}x \vert\det e\,\vert_{\vert x} \ \phi^{\dagger} \ (-i)^{\dagger}(e^{-1})^{\mu}_{\,\ c}(\gamma^{c})^{\dagger}\frac{1}{4}(-\omega^{\dagger}_{ba\mu})(\gamma^{a})^{\dagger}(\gamma^{b})^{\dagger} \ \psi} \\
\nnum &=& \int{d^{2}x \vert\det e\,\vert_{\vert x} \left(i(e^{-1})^{\mu}_{\,\ c}\frac{1}{4}\omega_{ba\mu}\gamma^{b}\gamma^{a} \gamma^{c} \phi\right)^{\dagger}\psi}
\eeq
and the following relation holds : $\gamma^{b}\gamma^{a}\gamma^{c} = \gamma^{c}\gamma^{b}\gamma^{a} + 2\ \delta^{ac}\ \gamma^{b} - 2\ \delta^{bc}\ \gamma^{a} $\\
thus
\beq
(II) &=& \int{d^{2}x \vert\det e\,\vert_{\vert x} \left(i(e^{-1})^{\mu}_{\,\ c}\frac{1}{4}(\omega_{ba\mu}\gamma^{c}\gamma^{b}\gamma^{a} +2\ {{\omega_{b}}^{c}}_{\mu}\gamma^{b} -2\ {\omega^{c}}_{a\mu}\gamma^{a}) \phi\right)^{\dagger}\psi} \\
\nnum &=& \int{d^{2}x \vert\det e\,\vert_{\vert x} \left(i(e^{-1})^{\mu}_{\,\ c}\gamma^{c}\frac{1}{4}\omega_{ab\mu}\gamma^{a} \gamma^{b} \phi\right)^{\dagger}\psi} +\int{d^{2}x \vert\det e\,\vert_{\vert x} \left(i{(e^{-1})^{\mu}}_{a}{{\omega_{b}}^{a}}_{\mu}\gamma^{b} \phi\right)^{\dagger}\psi}
\eeq
\noi The second term reads:
\beq
(II) &=& \int{d^{2}x \vert\det e\,\vert_{\vert x} \left(i(e^{-1})^{\mu}_{\,\ c}\gamma^{c}\frac{1}{4}\omega_{ab\mu}\gamma^{a} \gamma^{b} \phi\right)^{\dagger}\psi} \\
\nonumber &\ & +i \int{d^{2}x \vert\det e\,\vert_{\vert x} \ (e^{-1})^{\mu}_{\,\ a}\omega^{a}_{\,\ b\mu} \  \phi^{\dagger}\gamma^{b}\psi}
\eeq
\\

\noi $\bullet$ One obtains:
\beq
\nnum \left( \phi,D\psi\right)&=& (I)+(II) \\
\nnum &=& \int{d^{2}x \vert\det e\,\vert_{\vert x} \left(i (e^{-1})^{\mu}_{\,\ c} \gamma^{c} \frac{\partial \phi}{\partial x^{\mu}} \right)^{\dagger} \psi}+\int{d^{2}x \vert\det e\,\vert_{\vert x} \left(i(e^{-1})^{\mu}_{\,\ c}\gamma^{c}\frac{1}{4}\omega_{ab\mu}\gamma^{a} \gamma^{b} \phi\right)^{\dagger}\psi} \\
&\ & -i\int{d^{2}x \frac{\partial}{\partial x^{\mu}}\left[ \vert\det e\,\vert_{\vert x} (e^{-1})^{\mu}_{\,\ c} \right] \phi^{\dagger}\gamma^{c}\psi} +i \int{d^{2}x \vert\det e\,\vert_{\vert x} (e^{-1})^{\mu}_{\,\ a}\omega^{a}_{\,\ b\mu} \phi^{\dagger}\gamma^{b}\psi} \\
\nnum &=& \left( D\phi,\psi\right) + i\int{d^{2}x \vert\det e\,\vert_{\vert x} \left((e^{-1})^{\mu}_{\,\ a}\omega^{a}_{\,\ b\mu}-\frac{1}{\vert\det e\,\vert_{\vert x}}\frac{\partial}{\partial x^{\mu}}\left[ \vert\det e\,\vert_{\vert x} (e^{-1})^{\mu}_{\,\ c} \right]\right) \phi^{\dagger}\gamma^{b}\psi} \\
\nnum &=& \left( D\phi,\psi\right) + i \int{d^{2}x \vert\det e\,\vert_{\vert x} \ F_{b} \  \phi^{\dagger}\gamma^{b}\psi}
\eeq

\beq
\textrm{où} \ F_{b}=(e^{-1})^{\mu}_{\,\ a}\omega^{a}_{\,\ b\mu}-\frac{1}{\vert\det e\,\vert_{\vert x}}\frac{\partial}{\partial x^{\mu}}\left[ \vert\det e\,\vert_{\vert x} (e^{-1})^{\mu}_{\,\ c} \right]
\eeq
\noi Using the relations:
\beq
\frac{\partial (e^{-1})^{\mu}_{\,\ a}}{\partial x^{\sigma}} &=& - (e^{-1})^{\mu}_{\,\ b}\frac{\partial e^{b}_{\,\ \rho}}{\partial x^{\sigma}}(e^{-1})^{\rho}_{\,\ a} \\
\frac{\partial \ \det A}{\partial x^{\mu}} &=& \det A \cdot\ Tr\left[A^{-1}\frac{\partial A}{\partial x^{\mu}} \right] = \det A \cdot\ (A^{-1})^{\rho}_{\,\ a}\frac{\partial A^{a}_{\,\ \rho}}{\partial x^{\mu}}
\eeq
\noi this lead to the calculations:
\beq
F_{b}&=&(e^{-1})^{\mu}_{\,\ a}\omega^{a}_{\,\ b\mu}-\frac{1}{\vert\det e\,\vert_{\vert x}}\frac{\partial}{\partial x^{\mu}}\left[ \vert\det e\,\vert_{\vert x} (e^{-1})^{\mu}_{\,\ c} \right] \\
\nnum &=&(e^{-1})^{\mu}_{\,\ a}\omega^{a}_{\,\ b\mu} - \frac{1}{\vert\det e\,\vert_{\vert x}}\frac{\partial}{\partial x^{\mu}}\left[ \vert\det e\,\vert_{\vert x} \right] (e^{-1})^{\mu}_{\,\ c} - \frac{\partial}{\partial x^{\mu}}\left[ (e^{-1})^{\mu}_{\,\ c} \right] \\
\nnum &=& (e^{-1})^{\mu}_{\,\ a}\omega^{a}_{\,\ b\mu}- (e^{-1})^{\sigma}_{\,\ a}\frac{\partial e^{a}_{\,\ \sigma}}{\partial x^{\mu}}(e^{-1})^{\mu}_{\,\ c} + (e^{-1})^{\mu}_{\,\ a}\frac{\partial e^{a}_{\,\ \nu}}{\partial x^{\mu}}(e^{-1})^{\nu}_{\,\ b} \\
\nnum &=& (e^{-1})^{\mu}_{\,\ a}\omega^{a}_{\,\ b\mu} + (e^{-1})^{\mu}_{\,\ a} \left( \frac{\partial e^{a}_{\,\ \nu}}{\partial x^{\mu}} - \frac{\partial e^{a}_{\,\ \mu}}{\partial x^{\nu}} \right)(e^{-1})^{\nu}_{\,\ b}
\eeq
\noi Moreover, the torsion 2-form reads ($1^{st}$ structure equation):
\beq
T&:=& de + \omega\wedge e \\
\nnum &=& \frac{1}{2} \left[ (\partial_{\mu}e^{a}_{\,\ \nu}-\partial_{\nu}e^{a}_{\,\ \mu}) + (\omega^{a}_{\,\ b\mu}e^{b}_{\,\ \nu} - \omega^{a}_{\,\ b\nu}e^{b}_{\,\ \mu}) \right] dx^{\mu} \wedge dx^{\nu} \\
\nnum &=& \frac{1}{2} \ T^{a}_{\,\ \mu \nu} \ dx^{\mu} \wedge dx^{\nu}
\eeq

\beq
\Longrightarrow \partial_{\mu}e^{a}_{\,\ \nu}-\partial_{\nu}e^{a}_{\,\ \mu} = T^{a}_{\,\ \mu \nu} - (\omega^{a}_{\,\ b\mu}e^{b}_{\,\ \nu} - \omega^{a}_{\,\ b\nu}e^{b}_{\,\ \mu})
\eeq
\noi and thus
\beq
F_{b}&=& (e^{-1})^{\mu}_{\,\ a}\omega^{a}_{\,\ b\mu} + (e^{-1})^{\mu}_{\,\ a} \left[T^{a}_{\,\ \mu \nu} - (\omega^{a}_{\,\ c\mu}e^{c}_{\,\ \nu} - \omega^{a}_{\,\ c\nu}e^{c}_{\,\ \mu}) \right](e^{-1})^{\nu}_{\,\ b} \\
\nnum &=& (e^{-1})^{\mu}_{\,\ a}\omega^{a}_{\,\ b\mu} + T^{a}_{\,\ \mu \nu} (e^{-1})^{\mu}_{\,\ a} (e^{-1})^{\nu}_{\,\ b} - (e^{-1})^{\mu}_{\,\ a}\omega^{a}_{\,\ b\mu} + (e^{-1})^{\mu}_{\,\ b}\omega^{a}_{\,\ a\mu} \\
\nnum &=& T^{a}_{\,\ ab}
\eeq
\noindent where the torsion $T$ is function of the Lévy-Civita connection $\Gamma^{\alpha}_{\,\ \mu \nu}$:
\beq
T^{a}_{\,\ ab} &=& e^{a}_{\,\ \alpha} \ T^{\alpha}_{\,\ \mu \nu} \ (e^{-1})^{\mu}_{\,\ a} (e^{-1})^{\nu}_{\,\ b} \\
\nnum &=& e^{a}_{\,\ \alpha}\left[ \Gamma^{\alpha}_{\,\ \mu \nu} - \Gamma^{\alpha}_{\,\ \nu \mu}
\right](e^{-1})^{\mu}_{\,\ a} (e^{-1})^{\nu}_{\,\ b}
\eeq
\noi One finally obtains:
\beq
\left( \phi,D\psi\right)&=& \left( D\phi,\psi\right) + i\int{d^{2}x \vert\det e\,\vert_{\vert x} \ T^{a}_{\,\ ab} \ \phi^{\dagger}\gamma^{b}\psi}
\eeq
Hence, hermiticity of the Dirac operator implies the constraint:
\beq
\label{Contr}T^{a}_{\,\ ab} = 0
\eeq
or in a similar way:
\beq
T^{a}_{\,\ a\mu} = T^{a}_{\,\ ab}\ {e^{b}}_{\mu} = 0
\eeq

However, in dimension 2, this constraint implies that torsion itself vanishes. We show this property in the following. \\

The torsion tensor ${T^{a}}_{\mu\nu}$ (which can be qualified of hybrid because its covariants and contravariants indices are expressed in different frames) is an antisymmetric tensor on its two covariant indices:
\beq
{T^{a}}_{\mu\nu} = -\ {T^{a}}_{\nu\mu}
\eeq
and thus possesses, in dimension $n$, $\frac{n^2(n-1)}{2}$ independent components. In dimension 2, it possesses two degrees of freedom. The previous constraint then writes:
\beq
{T^{a}}_{a\mu} &=& {\left(e^{-1}\right)^{\sigma}}_{a}\ {T^{a}}_{\sigma\mu} \\
\nnum &=& \left\{\bar{lll} {\left(e^{-1}\right)^{2}}_{1}\ {T^{1}}_{21}+{\left(e^{-1}\right)^{2}}_{2}\ {T^{2}}_{21} = 0 &\ \ & (\mu\,=\,1) \\ {\left(e^{-1}\right)^{1}}_{1}\ {T^{1}}_{12}+{\left(e^{-1}\right)^{1}}_{2}\ {T^{2}}_{12} = 0 &\ \ & (\mu\,=\,2) \ear\right.
\eeq
that is equivalent to the system:
\beq
\left\{\bar{c} {\left(e^{-1}\right)^{2}}_{1}\ {T^{1}}_{12} = {\left(e^{-1}\right)^{2}}_{2}\ {T^{2}}_{21} \\ {\left(e^{-1}\right)^{1}}_{1}\ {T^{1}}_{12} = {\left(e^{-1}\right)^{1}}_{2}\ {T^{2}}_{21}\ear\right.
\eeq
Making the ratio of these two equations, one obtains the constraint:
\beq
\frac{{\left(e^{-1}\right)^{2}}_{1}}{{\left(e^{-1}\right)^{1}}_{1}} = \frac{{\left(e^{-1}\right)^{2}}_{2}}{{\left(e^{-1}\right)^{1}}_{2}}\ \ \ &\Longleftrightarrow&\ \ \ {\left(e^{-1}\right)^{1}}_{1}\ {\left(e^{-1}\right)^{2}}_{2} - {\left(e^{-1}\right)^{2}}_{1}\ {\left(e^{-1}\right)^{1}}_{2} = 0 \\
\nnum &\Longleftrightarrow&\ \ \ \det e^{-1} = 0\ \ \ \ \textrm{impossible :}\ \ {e^{a}}_{\mu}\in GL_{4}(\RR)
\eeq
\beq
\hspace{-2cm}\Longrightarrow\ \ \ {T^{1}}_{12} = {T^{2}}_{12} = 0
\eeq
Hence, in dimension 2, hermiticity of the Dirac operator implies that space cannot have torsion.

\newpage
\section{1$^{\textrm{st}}$ coordinate~transformation~:~cartesian~/ polar}
\markright{Coordinate~transformation~:~cartesian~/ polar}
To realize this first fluctuation, we consider the case of a flat euclidian space of dimension 2 without torsion, on which a coordinate system ${\tilde x}^{\mu} = \left( x, y \right)^{T}$ is defined leading to the holonomic and orthonormal frame such that:
\beq
\tilde\pa_{\mu}\ &:=& \left\{\frac{\pa}{\pa {\tilde x}^{\mu}}\right\}_{\mu\, =\, 1}^{2} = \left\{ \frac{\pa}{\pa x}, \frac{\pa}{\pa y} \right\} \\
\nnum \textrm{et}\ \ \ \ \tilde g_{\mu\nu} &=& \delta_{\mu\nu}
\eeq
\noi In this frame, the Dirac operator reads:
\beq
\tilde\Ds_{\vert\tilde x} = i{\left(\tilde e^{-1}(\tilde x)\right)^{\mu}}_{c}\ \gamma^{c} \left(\frac{\pa}{\pa \tilde x^{\mu}}+\frac{1}{4}
\left(\tilde\omega(\tilde x)\right)_{ab\mu}\gamma^{a}\gamma^{b}\right)
\eeq
\noi The vierbein ${{\tilde e}^{a}}_{\ \mu}$, chosen in the symmetric gauge and defined as the square root of the metric tensor, corresponds to the unit matrix. Moreover, this space being without torsion, the spin connection $\tilde\omega(\tilde x)$ vanishes identically. The Dirac operator $\tilde\Ds$  reduces to:
\beq
\tilde\Ds = \tilde\ds =\ i\ {\tilde\delta^{\mu}}_{\ a}\ \gamma^{a} \frac{\pa}{\pa \tilde x^{\mu}}
\eeq
\noi Now, one considers  polar coordinates $x^{\mu} = \varphi\left( \tilde x \right)^{\mu} = \left( r,\theta \right)^{T}$, in the holonome frame:
\beq
\left\{\bar{l}r\ \in\ \left[0,+\infty\right] \\ \theta\ \in\ \left[0,2\pi\right[\ear\right.
\eeq
\vspace{-1cm}
\beq
\pa_{\mu} := \left\{\frac{\pa}{\pa x^{\mu}}\right\}_{\mu\, =\, 1}^{2} = \left\{ \frac{\pa}{\pa r}, \frac{\pa}{\pa \theta} \right\}\ \ \ \ \textrm{endowed with the metric tensor}\ \ \ \ (g(x))_{\mu\nu}
\eeq
\noi which is deduced from ${\tilde x}^{\mu}$ by the diffeomorphism $\varphi$ such that:
\beq
{\tilde x}^{\mu} &=& \left(\varphi^{-1}(x)\right)^{\mu} \Longleftrightarrow \left\{\bar{l}
x = r \cos(\theta) \\ y = r \sin(\theta)\ear\right. \\
\nnum d{\tilde x}^{\mu} &=& \frac{\pa \left(\varphi^{-1}(x)\right)^{\mu}}{\pa x^{\nu}}\ dx^{\nu} = {\left(J_{\svphi}^{-1}(x)\right)^{\mu}}_{\nu}\ dx^{\nu} \\
\nnum \textrm{where}\ \ \ \ {\left(J_{\svphi}^{-1}(x)\right)^{\mu}}_{\nu} &=&
\left(\bar{cc}
\cos \left(\theta\right) & - r \sin(\theta)\\
\sin(\theta)  & r \cos(\theta)
\ear\right)
\eeq

\noi Now, we have to determine the resulting Dirac operator $\Ds :={^{\varphi}\!\Ds}$ such that:
\beq
{\Ds} &=& L\left(\varphi\right)\tilde\Ds\ L\left(\varphi\right)^{-1} \\
\nnum &=& i{\left(e^{-1}(x)\right)^{\mu}}_{c}\ \gamma^{c} \left(\frac{\pa}{\pa x^{\mu}}+\frac{1}{4}
\left(\omega(x)\right)_{ab\mu}\gamma^{a}\gamma^{b}\right)
\eeq
We thus have to determine the expression of the vierbein ${\left(e(x)\right)^{a}}_{\mu}$ and of the spin connection $\left(\omega(x)\right)_{ab\mu}$. \\

\noi From the previous jacobian, we can compute the metric tensor $(g(x))_{\mu\nu}$ :
\beq
d{\tilde x}^{\alpha}\ d{\tilde x}_{\alpha} &=& \tilde g_{\mu\nu}\ d{\tilde x}^{\mu}\ d{\tilde x}^{\nu} \\
\nnum &=& \tilde g_{\rho\sigma}\ {\left(J_{\svphi}^{-1}(x)\right)^{\rho}}_{\mu}\ {\left(J_{\svphi}^{-1}(x)\right)^{\sigma}}_{\nu}\ dx^{\mu}\ dx^{\nu} \\
\nnum &=& \left(g(x)\right)_{\mu\nu}\ dx^{\mu}\ dx^{\nu} \\
\nnum \Longrightarrow\ \  \left(g(x)\right)_{\mu\nu} &=& \left(J_{\svphi}^{-T}(x)\ \tilde g\ J_{\svphi}^{-1}(x)\right)_{\mu\nu}
\eeq
\noi One then obtains :
\beq
\left(g(x)\right)_{\mu\nu} = g\left( \frac{\pa}{\pa x^{\mu}},\frac{\pa}{\pa x^{\nu}} \right)_{\vert x} =
\left(\bar{cc}
1 & 0 \\ 0 & r^{2}
\ear\right)
\eeq
One can observe that the basis $\pa_{\mu}$ is not normed for this metric. \\
\noi We will find the vierbein allowing one to obtain an orthonormal frame $\left\{ e_{a} \right\}$ for this metric.
We write $\left\{ e^{a} \right\}$ the dual orthonormal basis: $e^{a} = {\left(e(x)\right)^{a}}_{\mu}\ dx^{\mu}$.
\beq
e_{a} = {\left(e^{-1}(x)\right)^{\mu}}_{a}\ \frac{\pa}{\pa x^{\mu}}
\eeq
\noi Since this base is orthonormal, the following relation holds:
\beq
\left(g(x)\right)_{ab} = g\left( e_{a}, e_{b} \right)_{\vert x} = \left(g(x)\right)_{\mu\nu}\ {\left(e^{-1}(x)\right)^{\mu}}_{a}\ {\left(e^{-1}(x)\right)^{\nu}}_{b} = \delta_{ab}
\eeq
\vspace{-0.8cm}
\beq
\noi \Longrightarrow\ \ \left(e^{-T}ge^{-1}\right)_{ab} = \delta_{ab}
\eeq

\noi We have chosen to express vierbein matrices ${\left(^{\varphi_{i}}e(x)\right)^{a}}_{\mu}$ in the symmetric gauge. The metric tensor $\left(g(x)\right)_{\mu\nu}$ being diagonal, the computation of its square root is straightforward:
\beq
{\left(e(x)\right)^{a}}_{\mu} &=& {\left(\sqrt{g(x)}\ \right)^{a}}_{\mu} =
\left(\bar{cc}
1 & 0 \\ 0 & r
\ear\right) \\
\nnum\Longrightarrow\ \ {\left(e^{-1}(x)\right)^{\mu}}_{a} &=&
\left(\bar{cc}
1 & 0 \\ 0 & 1/r
\ear\right)
\eeq

\noi The (dual) frame $\left\{e^{a}\right\}$ is orthonormal but no more holonomic:
\beq
\left\{\bar{l} e^{1} = dr \\ e^{2} = r d\theta \ear\right. \Longrightarrow \left\{\bar{l} de^{1} = 0 \\ de^{2} = \frac{e^{1}\wedge e^{2}}{r} = dr\wedge d\theta \ear\right.
\eeq

\noi We compute the spin connection $\omega(x)$ using the following formula, obtained from solving the first Cartan structure equation in the case of a space without torsion (and thus restrained to this particular case):
\beq
\label{Connection}\omega\wedge e &=& -\,de\ \ \ \ \Longrightarrow \\
\nnum {\left(\omega(x)\right)^{a}}_{b\mu} &=& \frac{1}{2} \left[\pa_{\beta}{e^{a}}_{\mu} - \pa_{\mu}{e^{a}}_{\beta} + {e^{m}}_{\mu}\left(\pa_{\beta}{e^{m}}_{\alpha}\right){\left(e^{-1}\right)^{\alpha}}_{a}\right]{\left(e^{-1}\right)^{\beta}}_{b}\ -\ \left[ a\longleftrightarrow b \right]
\eeq
One then obtains :
\beq
\omega(x) := {\left(\omega(x)\right)^{a}}_{b} = \left(\bar{cc} 0 & -1 \\ 1 & 0 \ear\right)\ d\theta
\eeq
We can now verify that the frame $\pa_{\mu}$, the torsion 2-form $T(x)$ and the curvature scalar deduced from the curvature 2-form $R(x)$ (1$^{st}$ and 2$^{nd}$ Cartan structure equations) are null (in the case of the torsion, it is equivalent to verifying the computations done, this condition being the starting hypothesis for the computation of the connection):
\beq
T &=& de + \omega \wedge e \\
\nnum&&\left.\bar{l}
de = \pa_{\mu}{e^{a}}_{\nu}\ dx^{\mu}\wedge dx^{\nu} = \left(\bar{c} 0 \\ 1 \ear\right)\ dr\wedge d\theta \\
\omega\wedge e = {\omega^{a}}_{b\mu}\ {e^{b}}_{\nu}\ dx^{\mu}\wedge dx^{\nu} \\
\ \ \ \ \ \ \ \, = \left(\bar{cc} 0 & -1 \\ 1 & 0 \ear\right)
\left(\bar{l} 1 \\ 0 \ear\right)\ d\theta\wedge dr = \left(\bar{c} 0 \\ -1 \ear\right)\ d\theta\wedge dr
\ear \right\} \ \Longrightarrow\ \ T = 0 \\
R &= &d\omega + \omega\wedge\omega \\
\nnum&&\left.\bar{l}
d\omega = \pa_{\mu}{\omega^{a}}_{b\nu}\ dx^{\mu}\wedge dx^{\nu} = 0 \\
\omega\wedge\omega = {\omega^{a}}_{b\mu}\ {\omega^{b}}_{c\nu}\ dx^{\mu}\wedge dx^{\nu} = 0 \\
\ear\right\}\ \ \ \Longrightarrow\ \ \ R = 0
\eeq
Hence, starting with a flat, torsionless space, we obtain, after applying the diffeomorphism, a still flat and torsionless space. \\

The fluctuation is operated in the following. In order to simplify the calculations, the $\alpha_{i}$ coefficients are taken equal to unity. The fluctuation will be done in the frame deriving from the polar coordinate system. Hence, the diffeomorphism mapping cartesian coordinates to
 the polar ones is $\varphi^{-1}$. The jacobian that appears in formula (\ref{Fluc006}) is $J_{\svphi}$ associated to ${\left(\tilde e^{-1}\right)^{a}}_{ \mu}$ :
\beq
\label{App1Eq1}\nnum {\left(E^{-1}(x)\right)^{\mu}}_{c} &:=& \vert\det\ef\,\vert_{\vert x}\ {\left(\ef^{-1}(x)\right)^{\mu}}_{c} \\
&=& \vert\det e\,\vert_{\vert x}\ {\left[J_{\svphi}(x)\ {\tilde e^{-1}\left(\varphi^{-1}(x)\right)}+e^{-1}(x)\right]^{\mu}}_{c} \\
\nnum &=& r\ \left[ \left(\bar{cc} \cos(\theta) & \sin(\theta) \\ -\sin(\theta)/r & \cos(\theta)/r \ear\right)\ \id_{2} + \left(\bar{cc} 1 & 0 \\ 0 & 1/r \ear\right)\right] \\
\nnum &=& \left(\bar{cc} r\left(\cos(\theta)+1\right) & r\sin(\theta) \\ -\sin(\theta) & \left(\cos(\theta)+1\right) \ear\right)
\eeq
We also compute the term $\left(\Omega_{E}(x)\right)_{abc}$ :
\beq
\label{App1Eq2}\nnum \left(\Omega_{E}(x)\right)_{abc} &:=& \vert\det\ef\,\vert_{\vert x} {\left(\ef^{-1}(x)\right)^{\mu}}_{c} \left(\omegaf(x)\right)_{ab\mu} \\
\nnum &=& \vert\det e\,\vert_{\vert x}\ {\left(e^{-1}(x)\right)^{\mu}}_{c}\ \left(\omega(x)\right)_{ab\mu} \\
\nnum &=& \vert\det e\,\vert_{\vert x}\ \left(\omega(x)\right)_{ab}\ {\left(e^{-1}(x)\right)}_{c} \\
&=& r\ \left(\bar{ccc} 0_{2_{(ab)}} &,& \left(\bar{cc} 0 & -1 \\ 1 & 0 \ear\right)_{\!\!\!\!(ab)} \ear \right)_{\!\!\!\!(\mu)} \left(\bar{cc} 1 & 0 \\ 0 & 1/r \ear\right)_{\!\!\!\!(\mu c)} \\
\nnum &=& \left(\bar{ccc} 0_{2_{(ab)}} &,& \left(\bar{cc} 0 & -1 \\ 1 & 0 \ear\right)_{\!\!\!\!(ab)} \ear \right)_{\!\!\!\!(c)}
\eeq
where bracketed indices mark which matrix is related to the same un-bracketed indices.\\
It would be easy to express, by means of these terms, the vierbein and the spin connection of the fluctuated Dirac operator. However, in order to understand each step of the calculation, we forget these terms and pursue with the same method as the previous one, and keep the use of the general terms $E(x)$ and $\Omega_{E}(x)$ to the next example which is more expensive in calculations. \\
\noi Thus, we compute the determinant of the fluctuated vierbein :
\beq
\textrm{the space being of dimension 2,}\ \det\left[(\ref{App1Eq1})\right]\ \longrightarrow\ \ \left(\det\ef\,\right)_{\vert x}
\eeq
\vspace{-0.8cm}
\beq
\vert\det\ef\,\vert_{\vert x} = r \left[\left(\cos(\theta)+1\right)^{2} + \sin^{2}(\theta)\right] = 2r \left(\cos(\theta)+1\right)
\eeq
Recalling that the first two lines of (\ref{App1Eq1}) and (\ref{App1Eq2}) are obtained by identification, we compute, thanks to (\ref{App1Eq1}), the vierbein ${\left(\ef(x)\right)^{a}}_{\mu}$:
\beq
\nnum {\left(\ef^{-1}(x)\right)^{\mu}}_{c} &=& \frac{1}{\vert\det\ef\,\vert_{\vert x}}\ \vert\det e\,\vert_{\vert x}\ {\left[J_{\svphi}(x)\ \vert\det e\,\vert_{\vert x}{\tilde e^{-1}\left(\varphi^{-1}(x)\right)}+e^{-1}(x)\right]^{\mu}}_{c} \\
&=& \frac{1}{2r \left(\cos(\theta)+1\right)}\ \left(\bar{cc} r\left(\cos(\theta)+1\right) & r\sin(\theta) \\
-\sin(\theta) & \left(\cos(\theta)+1\right) \ear\right) \\
\nnum &=& \left(\bar{cc} 1/2 & \frac{\sin(\theta)}{2 \left(\cos(\theta)+1\right)} \\
\frac{-\sin(\theta)}{2r \left(\cos(\theta)+1\right)} & 1/2r \ear\right)
\eeq
\beq
{\left(\ef(x)\right)^{c}}_{\mu} &=& \left(\bar{cc} \cos(\theta)+1 & -r\sin(\theta) \\
\sin(\theta) & r\left(\cos(\theta)+1\right) \ear\right)
\eeq
We see that we are no more in the symmetric gauge. \\
Using the fluctuated vierbein and its determinant, we compute by identification, thanks to (\ref{App1Eq2}), the fluctuated spin connection $\left(\omegaf(x)\right)_{ab\mu}$:
\beq
\left(\omegaf(x)\right)_{ab\mu} &=& \frac{1}{\vert\det\ef\,\vert_{\vert x}}\ \left( \vert\det e\,\vert_{\vert x}\ {\left(e^{-1}(x)\right)^{\rho}}_{c}\ \left(\omega(x)\right)_{ab\rho}\right)
{\left(\ef(x)\right)^{c}}_{\mu}\\
\nnum &=& \frac{1}{2r \left(\cos(\theta)+1\right)}\ \left(\bar{ccc} 0_{2_{(ab)}} &,& \left(\bar{cc} 0 & -1 \\
1 & 0
\ear\right)_{\!\!\!\!(ab)} \ear \right)_{\!\!\!\!(c)}\ \left(\bar{cc} \cos(\theta)+1 & -r\sin(\theta) \\
\sin(\theta) & r\left(\cos(\theta)+1\right) \ear\right)_{\!\!\!\!(c\mu)} \\
\nnum &=& \left(\bar{ccc} \frac{\sin(\theta)}{2r \left(\cos(\theta)+1\right)}\ \left(\bar{cc} 0 & -1 \\
1 & 0
\ear\right)_{\!\!\!\!(ab)} &,& \frac{1}{2}\ \left(\bar{cc} 0 & -1 \\
1 & 0
\ear\right)_{\!\!\!\!(ab)} \ear \right)_{\!\!\!\!(\mu)} \\
\nnum &=& \left(\bar{cc} 0 & -1 \\
1 & 0
\ear\right)_{\!\!\!\!(ab)}\ \left(\bar{ccc} \frac{\sin(\theta)}{2r \left(\cos(\theta)+1\right)}\ &,& \frac{1}{2}\ear \right)_{\!\!\!(\mu)}
\eeq
Since latin indices belong to tetrads, in which the metric tensor is $\eta_{\mu\nu}$, and since we consider an euclidian space ($\eta_{\mu\nu} = \delta_{\mu\nu}$), it follows:
\beq
{\left(\omegaf(x)\right)^{a}}_{b\mu} = \delta^{ac}\ \left(\omegaf(x)\right)_{cb\mu} = \left(\bar{cc} 0 & -1 \\
1 & 0
\ear\right)_{\!\!\!\!(ab)}\ \left(\bar{ccc} \frac{\sin(\theta)}{2r \left(\cos(\theta)+1\right)}\ &,& \frac{1}{2}\ear \right)_{\!\!\!(\mu)}
\eeq
We have now to check hermiticity of the resulting Dirac operator. We have shown in the previous section that this property in a 2-dimensional space leads to ${^{f}T(x)} = 0$. This calculation follows:
\beq
d\ef &=& \pa_{\mu}{\ef^{a}}_{\nu}\ dx^{\mu}\wedge dx^{\nu} = \left(\pa_{1}{\ef^{a}}_{2}-\pa_{2}{\ef^{a}}_{1}\right)\ dx^{1}\wedge
dx^{2} \\
\nnum &=& \left(\bar{c} -\sin(\theta)+\sin(\theta) \\ \left(\cos(\theta)+1\right)-\cos(\theta) \ear\right)\ dr\wedge d\theta = \left(\bar{c} 0 \\ 1 \ear\right)\ dr\wedge d\theta
\eeq
\beq
\omegaf\wedge\ef &=& {\omegaf^{a}}_{b\mu}\ {\ef^{b}}_{\nu}\ dx^{\mu}\wedge dx^{\nu} = \left({\omegaf^{a}}_{b1}\
{\ef^{b}}_{2} - {\omegaf^{a}}_{b2}\ {\ef^{b}}_{1}\right)\ dx^{1}\wedge dx^{2} \\
\nnum &=& \left[\frac{\sin(\theta)}{2r\left(\cos(\theta)+1\right)} \left(\bar{cc} 0 & -1 \\
1 & 0
\ear\right) \left(\bar{c}-r\sin(\theta) \\
r\left(\cos(\theta)+1\right)\ear\right)\right. {}\\
\nnum &&{} \ \ \ \ \ - \left.\frac{1}{2} \left(\bar{cc} 0 & -1 \\
1 & 0
\ear\right) \left(\bar{c}\cos(\theta)+1 \\
\sin(\theta)\ear\right)\right]\ dr\wedge d\theta \\
\nnum &=& \left(\bar{c} 0 \\ -1 \ear\right)\ dr\wedge d\theta
\eeq
\beq
{^{f}T(x)} =\ \ d\ef\ +\ \omegaf\wedge\ef\ =\ 0
\eeq
The fluctuated Dirac operator satisfies the hermiticity condition.\\

What about the curvature of the resulting space. Let us first compute the curvature 2-form ${^{f}\!R}(x)$:
\beq
d\omegaf &=& \pa_{\mu}{\omegaf^{a}}_{b\nu}\ dx^{\mu}\wedge dx^{\nu} = \left(\pa_{1}{\omegaf^{a}}_{b2}-\pa_{2}{\omegaf^{a}}_{b1}\right)\ dx^{1}\wedge
dx^{2} \\
\nnum &=& -\ \pa_{2}{\omegaf^{a}}_{b1}\ dx^{1}\wedge dx^{2} \\
\nnum &=& \frac{-1}{2r \left(\cos(\theta)+1\right)} \left(\bar{cc} 0 & -1 \\
1 & 0
\ear\right)\ dx^{1}\wedge dx^{2}
\eeq
\beq
\omegaf\wedge\omegaf &=& {\omegaf^{a}}_{b\mu}\ {\omegaf^{b}}_{c\nu}\ dx^{\mu}\wedge dx^{\nu} = \left({\omegaf^{a}}_{b1}\ {\omegaf^{b}}_{c2} - {\omegaf^{a}}_{b2}\
{\omegaf^{b}}_{c1}\right)\ dx^{1}\wedge dx^{2} \\
\nnum &=& 0
\eeq
\beq
{^{f}\!R}(x) &=&\ d\omegaf\ +\ \omegaf\wedge\omegaf \\
\nnum &=& \frac{-1}{2r \left(\cos(\theta)+1\right)} \left(\bar{cc} 0 & -1 \\
1 & 0
\ear\right)\ dx^{1}\wedge dx^{2}
\eeq
In order to obtain the expression of the fluctuated scalar curvature ${^{f}\!R_{_{S}}}(x)$, we successively compute the Riemann tensor ${^{f}\!R(x)^{\alpha}}_{\beta\mu\nu}$, obtained by expressing every indice of the curvature 2-form in the same frame using the fluctuated vierbein, and the Ricci tensor ${^{f}\!R(x)}_{\mu\nu}$, obtained by contracting the first and third indices of the Riemann tensor.
\beq
{^{f}\!R}(x) &=& \frac{1}{2}\ {{^{f}\!R}^{a}}_{b\mu\nu}\ dx^{\mu}\wedge dx^{\nu} \\
\nnum &=& \frac{1}{2}\ {{^{f}\!R}^{\alpha}}_{\beta\mu\nu}\ {\left(\ef(x)\right)^{a}}_{\alpha}\ {\left(\ef^{-1}(x)\right)^{\beta}}_{b}\ dx^{\mu}\wedge
dx^{\nu}
\eeq
The Riemann tensor ${{^{f}\!R}^{\alpha}}_{\beta\mu\nu}$ is thus:
\beq
{{^{f}\!R}^{\alpha}}_{\beta1\,2}\ &=& {\left(\ef^{-1}(x)\right)^{\alpha}}_{a}\ {{^{f}\!R}^{a}}_{b1\,2}\
{\left(\ef(x)\right)^{b}}_{\beta} = -\ {{^{f}\!R}^{\alpha}}_{\beta\,21}\ \\
\nnum &=& \left(\bar{cc} 1/2 & \frac{\sin(\theta)}{2 \left(\cos(\theta)+1\right)} \\
\frac{-\sin(\theta)}{2r \left(\cos(\theta)+1\right)} & 1/2\,r \ear\right)\frac{-1}{r \left(\cos(\theta)+1\right)} \left(\bar{cc} 0 & -1 \\
1 & 0
\ear\right)\left(\bar{cc} \cos(\theta)+1 & -r\sin(\theta) \\
\sin(\theta) & r\left(\cos(\theta)+1\right) \ear\right)\\
\nnum &=& \left(\bar{cc} 0 & \frac{1}{\cos(\theta)+1} \\
\frac{-1}{r^{2} \left(\cos(\theta)+1\right)} & 0 \ear\right)
\eeq
Then follows the Ricci tensor:
\beq
{{^{f}\!R}}_{\mu\nu}\ &:=& {{^{f}\!R}^{\alpha}}_{\mu\alpha\nu}\ =
\left(\bar{cc} {{^{f}\!R}^{\alpha}}_{1\,\alpha1} & {{^{f}\!R}^{\alpha}}_{1\,\alpha2} \\
{{^{f}\!R}^{\alpha}}_{2\,\alpha1} & {{^{f}\!R}^{\alpha}}_{2\,\alpha2}\ear\right) = \left(\bar{cc} {{^{f}\!R}^{2}}_{1\,2\,1} & {{^{f}\!R}^{1}}_{1\,1\,2} \\
{{^{f}\!R}^{2}}_{2\,2\,1} & {{^{f}\!R}^{1}}_{2\,1\,2}\ear\right) \\
\nnum &=& \left(\bar{cc} \frac{1}{r^{2} \left(\cos(\theta)+1\right)} & 0 \\
0 & \frac{1}{\cos(\theta)+1}\ear\right)
\eeq
We finally obtain the curvature scalar :
\beq
{^{f}\!R_{_{S}}}(x) =\ {{^{f}\!R}}_{\mu\nu}\ {^{f}\!g}^{\mu\nu}
\eeq
Where we have used the usual convention : $\ g^{\mu\nu} :=
\left(g_{..}^{-1}\right)^{\mu\nu}$. Using the definition of the vierbein (and since we consider an euclidian space):
\beq
\left(e^{-T}\ g\ e^{-1}\right)_{\mu\nu} &=& \delta_{\mu\nu} \\
\nnum \Longrightarrow\ \ g^{\mu\nu}\ =\ \left(\left(e^{T}\ e\right)^{-1}\right)^{\mu\nu} &=& \left(e^{-1}\
e^{-T}\right)^{\mu\nu}
\eeq
We give, here after, the metric tensor corresponding to the fluctuated metric $^{f}\!g(x)$, expressed in the frame $\pa_{\mu}$:
\beq
\left({^{f}\!g}(x)\right)_{\mu\nu} &=& \left(\ef^{T}\ \ef\right)_{\mu\nu} =\ 2\left(\cos(\theta)+1\right) \left(\bar{cc}1&0\\0&r^{2}\ear\right) \\
\nnum &=& 2\left(\cos(\theta)+1\right)\ \left(g(x)\right)_{\mu\nu}
\eeq
The curvature scalar is finally given by:
\beq
{^{f}\!R_{_{S}}}(x) &=&\ {{^{f}\!R}}_{\mu\nu}\ {^{f}\!g}^{\mu\nu} \\
\nnum &=& \textrm{Tr}\left[\left(\bar{cc} \frac{1}{r^{2} \left(\cos(\theta)+1\right)} & 0 \\
0 & \frac{1}{\cos(\theta)+1}\ear\right)\left(\bar{cc} \frac{1}{2 \left(\cos(\theta)+1\right)} & 0 \\
0 & \frac{1}{2r^{2} \left(\cos(\theta)+1\right)}\ear\right)\right] \\
\nnum &=&\ \frac{1}{r^{2} \left(\cos(\theta)+1\right)^{2}}
\eeq
At the end of this first example, we have generated a space with non-vanishing curvature starting from a flat space. However, the dimension of space (dimension 2), together with the constraint (\ref{Contr}) related to the hermiticity of the Dirac operator, does not allow to generate torsion. The next example will allow to verify that this is however possible in the general case.

\newpage
\section{2$^{\textrm{nd}}$ coordinate~transformation~:~cartesian~/ spherical}
\markright{Coordinate~transformation~:~cartesian~/ spherical}
In this example, we consider an euclidian space, initially flat and without torsion, of dimension 3.
\noi We consider the cartesian coordinates ${\tilde x}^{\mu} = \left( x, y, z \right)^{T}$ in the holonomic and orthonormal frame :
\beq
\nnum \left\{x,y,z\right\}\ &\in& \left]-\infty, +\infty\right[ \\
\tilde\pa_{\mu}\ &:=& \left\{\frac{\pa}{\pa {\tilde x}^{\mu}}\right\}_{\mu\, =\, 1}^{3} = \left\{ \frac{\pa}{\pa x}, \frac{\pa}{\pa y}, \frac{\pa}{\pa z} \right\} \\
\nnum \textrm{et}\ \ \ \ \tilde g_{\mu\nu} &=& \delta_{\mu\nu}
\eeq
\noi In this frame, the Dirac operator is expressed as:
\beq
\tilde\Ds_{\vert\tilde x} = i{\left(\tilde e^{-1}(\tilde x)\right)^{\mu}}_{c}\ \gamma^{c} \left(\frac{\pa}{\pa \tilde x^{\mu}}+\frac{1}{4}
\left(\tilde\omega(\tilde x)\right)_{ab\mu}\gamma^{a}\gamma^{b}\right)
\eeq
\noi For the same reasons as in the previous example, the Dirac operator $\tilde\Ds$ reduces to:
\beq
\tilde\Ds = \tilde\ds =\ i\ {\tilde\delta^{\mu}}_{\ a}\ \gamma^{a} \frac{\pa}{\pa \tilde x^{\mu}}
\eeq
\noi We consider now the spherical coordinates $x^{\mu} = \sigma\left( \tilde x \right)^{\mu} = \left( r,\theta, \varphi \right)^{T}$, in the holonomic frame:
\beq
\left\{\bar{l}r\ \in \left[0,+\infty\right[ \\ \theta\ \in \left[0,\pi\right] \\ \varphi\ \in \left[0,2\pi\right[ \ear\right.
\eeq
\vspace{-0.8cm}
\beq
\nnum \pa_{\mu} := \left\{\frac{\pa}{\pa x^{\mu}}\right\}_{\mu\, =\, 1}^{3} = \left\{ \frac{\pa}{\pa r}, \frac{\pa}{\pa \theta},\frac{\pa}{\pa \varphi} \right\}\ \ \ \ \textrm{endowed with the metric tensor}\ \ \ \ (g(x))_{\mu\nu}
\eeq
\noi These coordinates are deduced from ${\tilde x}^{\mu}$ by using the diffeomorphism $\sigma$ defined by:
\beq
{\tilde x}^{\mu} &=& \left(\sigma^{-1}(x)\right)^{\mu} \Longleftrightarrow \left\{\bar{l}
x = r \sin(\theta)\, \cos(\varphi) \\ y = r \sin(\theta)\ \sin(\varphi) \\ z = r \cos(\theta)\ear\right. \\
\nnum d{x}^{\mu} &=& \frac{\pa \left(\sigma(\tilde x)\right)^{\mu}}{\pa \tilde x^{\nu}}\ d\tilde x^{\nu} = {\left(J_{\sigma}(\tilde x)\right)^{\mu}}_{\nu}\ d\tilde x^{\nu} \\
\nnum \textrm{où}\ \ \ \ {\left(J_{\sigma}^{-1}(\sigma^{-1}(x))\right)^{\mu}}_{\nu} &=&
\left(\bar{ccc}
\sin(\theta)\cos(\varphi) & r \cos(\theta)\cos(\varphi) & - r \sin(\theta)\sin(\varphi) \\
\sin(\theta)\sin(\varphi)  & r \cos(\theta)\sin(\varphi) & r \sin(\theta)\cos(\varphi) \\
\cos(\theta) & -r\sin(\theta) & 0 \ear\right)
\eeq
\noi We define also:
\beq
{\left(J_{\sigma^{-1}}^{-1}(\sigma^{-1}(x))\right)^{\mu}}_{\nu} :=
{\left(J_{\sigma}(\sigma^{-1}(x))\right)^{\mu}}_{\nu} = \left(\bar{ccc}
\sin(\theta)\cos(\varphi) & \sin(\theta)\sin(\varphi) & \cos(\theta) \\
\frac{\cos(\theta)\cos(\varphi)}{r}  & \frac{\cos(\theta)\sin(\varphi)}{r} & \frac{-\sin(\theta)}{r} \\
\frac{-\sin(\varphi)}{r\sin(\theta)} & \frac{\cos(\varphi)}{r\sin(\theta)} & 0 \ear\right)
\eeq
\noi From this jacobian, we determine the metric tensor $(g(x))_{\mu\nu}$ :
\beq
(g(x))_{\mu\nu} = \left(\bar{ccc}1&0&0\\0&r^{2}&0\\0&0&r^{2}\sin^{2}(\theta)\ear\right)
\eeq
Choosing the symmetric gauge to express the vierbein, we obtain:
\beq
{\left(e(x)\right)^{a}}_{\mu} &=& \left(\bar{ccc}1&0&0\\0&r&0\\0&0&r\sin(\theta)\ear\right)\ \ \ \textrm{with}\
\  \vert\det e\,\vert_{\vert x} = r^{2}\sin(\theta)
\eeq
We compute the connection thanks to the formula (\ref{Connection}), emphasizing once more that this formula is restreint to spaces without torsion, requirement fulfilled here:
\beq
\omega(x) := {\left(\omega(x)\right)^{a}}_{b} = \left(\bar{ccc} 0 & -1 & 0 \\ 1 & 0 & 0 \\ 0 & 0 & 0 \ear\right)\ d\theta\ +\ \left(\bar{ccc} 0 & 0 & -\sin(\theta) \\ 0 & 0 & -\cos(\theta) \\ \sin(\theta) & \cos(\theta) & 0 \ear\right)\ d\varphi
\eeq
We state:
\beq
J_{x} = \left(\bar{ccc} 0 & 0 & 0 \\ 0 & 0 & -1 \\ 0 & 1 & 0 \ear\right),\ J_{y} = \left(\bar{ccc} 0 & 0 & 1 \\ 0 & 0 & 0 \\ -1 & 0 & 0 \ear\right),\ J_{z} = \left(\bar{ccc} 0 & -1 & 0 \\ 1 & 0 & 0 \\ 0 & 0 & 0 \ear\right)
\eeq
which are representations of dimension 3 of the three generators of the group $SO(3)$. \\
The spin connection becomes:
\beq
\omega(x) =\ J_{z}\ d\theta\ +\ \left(\cos(\theta)\ J_{x}\ -\ \sin(\theta)\ J_{y}\right)\ d\varphi
\eeq
We will now verify that, as we expect, the resulting vierbein and spin connection describe a flat, torsionless space.
\beq
T &=& de + \omega \wedge e \\
\nnum&&\left.\bar{l}
de = \pa_{\mu}{e^{a}}_{\nu}\ dx^{\mu}\wedge dx^{\nu} \\
\ \ \ \ = \left(\bar{c} 0 \\ 1 \\ 0 \ear\right)\ dr\wedge d\theta + \left(\bar{c} 0 \\ 0 \\ r\cos(\theta) \ear\right)\ d\theta\wedge d\varphi + \left(\bar{c} 0 \\ 0 \\ -\sin(\theta) \ear\right)\ d\varphi\wedge dr \\
\omega\wedge e = {\omega^{a}}_{b\mu}\ {e^{b}}_{\nu}\ dx^{\mu}\wedge dx^{\nu} \\
\ \ \ \ \ \ \ \, = - {\omega^{a}}_{b\,2}\ {e^{b}}_{1}\ dx^{1}\wedge dx^{2}\ +\ \left({\omega^{a}}_{b\,2}\ {e^{b}}_{3}\ - {\omega^{a}}_{b\,3}\ {e^{b}}_{2}\right) dx^{2}\wedge dx^{3} {} \\
\hspace{3cm} +\ {\omega^{a}}_{b\,3}\ {e^{b}}_{1}\ dx^{3}\wedge dx^{1} \\
\ \ \ \ \ \ \ \, = \left(\bar{c} 0 \\ -1 \\ 0 \ear\right)\ dr\wedge d\theta + \left(\bar{c} 0 \\ 0 \\ -r\cos(\theta) \ear\right)\ d\theta\wedge d\varphi + \left(\bar{c} 0 \\ 0 \\ \sin(\theta) \ear\right)\ d\varphi\wedge dr \ear\right\} \\
\nnum &&\Longrightarrow\ \ T = 0
\eeq
\beq
\nnum \textrm{} \\
R &=& d\omega + \omega\wedge\omega \\
\nnum&&\left.\bar{l}
d\omega = \pa_{\mu}{\omega^{a}}_{b\nu}\ dx^{\mu}\wedge dx^{\nu} = \pa_{2}{\omega^{a}}_{b\,3}\ dx^{2}\wedge dx^{3} \\
\ \ \ \ \,= \left(-\sin(\theta)\ J_{x}\ -\ \cos(\theta)\ J_{y}\right)\ d\theta\wedge d\varphi \\
\omega\wedge\omega = {\omega^{a}}_{b\mu}\ {\omega^{b}}_{c\nu}\ dx^{\mu}\wedge dx^{\nu} \\
\ \ \ \ \ \ \ \ \,= \left[J_{z}\left(\cos(\theta)\ J_{x}\ -\ \sin(\theta)\ J_{y}\right)\ - \left(\cos(\theta)\ J_{x}\ -\ \sin(\theta)\ J_{y}\right)\ J_{z}\right]\ d\theta\wedge d\varphi  \\
\ \ \ \ \ \ \ \ \,= \left(\cos(\theta)\ \left[J_{z},J_{x}\right]\ +\ \sin(\theta)\ \left[J_{y},J_{z}\right]\right)\  d\theta\wedge d\varphi  \\
\ \ \ \ \ \ \ \ \,= \left(\cos(\theta)\ J_{y}\ +\ \sin(\theta)\ J_{x}\right)\  d\theta\wedge d\varphi  \\
\ear\right\} \\
\nnum &&\ \ \Longrightarrow\ \ R = 0
\eeq
Once checked, we can operate the fluctuation. As was the case in the previous example, we will not operate the fluctuation in the frame deriving from the cartesian coordinate system $\tilde\pa_{\mu}$, but in the one deriving from the spherical coordinate system $\pa_{\mu}$. The diffeomorphism that transforms the cartesian coordinates to the spherical ones is $\sigma^{-1}$. Once more, in order to simplify the calculations, we take the coefficients $\alpha_{i} = 1$ , with $i = \{1,2\}$. \\

The calculations implied in this example being substantial, a part of them have been made using the software Mathematica: the check for hermiticity of the fluctuated Dirac operator as well as the computation of the resulting curvature and torsion of space after fluctuation. \\
Moreover we use, in this example, the general terms $E(x)$ and $\Omega_{E}(x)$, what will constitute a validation of the general fluctuation method.

\noi We begin the fluctuation by computing the term ${\left(E(x)\right)^{a}}_{\mu}$:
\beq
{\left(E^{-1}(x) \right)^{\mu}}_{c} &=& \vert \det e\, \vert_{\vert x}\ \sum_{i\, =\, 1}^{2}\
\alpha_{i}\ {\left(J^{-1}_{\varphi_{i}}(x)\right)^{\mu}}_{\nu}\ {\left(^{\varphi_{i}\!}e^{-1}(\varphi_{i}(x))\right)^{\nu}}_{c} \\
\nnum &=& \vert \det e\, \vert_{\vert x}\ \left[{\left(J^{-1}_{Id}(x)\right)^{\mu}}_{\nu}\ {\left(^{Id\!}e^{-1}(x)\right)^{\nu}}_{c}\ +\ {\left(J^{-1}_{\sigma^{-1}}(\sigma^{-1}(x))\right)^{\mu}}_{\nu}\ {\left(^{\sigma^{-1}\!}e^{-1}(\sigma^{-1}(x))\right)^{\nu}}_{c} \right] \\
\nnum &=& \vert \det e\, \vert_{\vert x}\ \left[{\left(\delta\right)^{\mu}}_{\nu}\ {\left(e^{-1}(x)\right)^{\nu}}_{c}\ +\ {\left(J(\sigma^{-1}(x))\right)^{\mu}}_{\nu}\ {\left(\tilde e^{-1}(\sigma^{-1}(x))\right)^{\nu}}_{c} \right] \\
\nnum &=& r^{2}\sin(\theta)\ \left[\id_{3}\ \left(\bar{ccc}1&0&0\\0&\frac{1}{r}&0\\0&0&\frac{1}{r\sin(\theta)}\ear\right)\right. {}\\
\nnum &&{}\ \ \ \ \ \ \ \ \ \ \ \ \ \ \ \ \ \ \ \ \ \ \ \ \left.+\ \left(\bar{ccc}
\sin(\theta)\cos(\varphi) & \sin(\theta)\sin(\varphi) & \cos(\theta) \\
\frac{\cos(\theta)\cos(\varphi)}{r}  & \frac{\cos(\theta)\sin(\varphi)}{r} & \frac{-\sin(\theta)}{r} \\
\frac{-\sin(\varphi)}{r\sin(\theta)} & \frac{\cos(\varphi)}{r\sin(\theta)} & 0 \ear\right)\ \id_{3} \right] \\
\nnum &=& \left(\bar{ccc}
r^{2}\sin(\theta)\left(\sin(\theta)\cos(\varphi)+1\right) & r^{2}\sin^{2}(\theta)\sin(\varphi) & r^{2}\cos(\theta)\sin(\theta) \\
r\sin(\theta)\cos(\theta)\cos(\varphi)  & r\sin(\theta)\left(\cos(\theta)\sin(\varphi)+1\right) & -r\sin^{2}(\theta) \\
-r\sin(\varphi) & r\cos(\varphi) & r \ear\right)
\eeq
Recalling that the cosecant is defined by: $\csc(\theta) = 1/\sin(\theta)$, we get after inversion of the matrix:
\beq
{\left(E(x) \right)^{a}}_{\mu} &=& \left(\bar{ccc}
\frac{\csc(\theta)}{2r^{2}} & \frac{\csc(\theta)}{2r\left(\sec(\theta+\varphi)+\tan(\theta+\varphi)\right)} & -\frac{\cos(\theta)+\sin(\varphi)}{2r\left(\sin(\theta+\varphi)+1\right)} \\
-\frac{\csc(\theta)\cos(\theta+\varphi)}{2r^{2}\left(\sin(\theta+\varphi)+1\right)}  & \frac{\csc(\theta)}{2r} & \frac{\sin(\theta)+\cos(\varphi)}{2r\left(\sin(\theta+\varphi)+1\right)} \\
\frac{\csc(\theta)\left(\cos(\theta)+\sin(\varphi)\right)}{2r^{2}\left(\sin(\theta+\varphi)+1\right)} & -\frac{\csc(\theta)\left(\sin(\theta)+\cos(\varphi)\right)}{2r\left(\sin(\theta+\varphi)+1\right)} & \frac{1}{2r}
\ear\right)
\eeq
and its determinant:
\beq
\left\vert\,\det E\,\right\vert_{\vert x} &=& \left[2r^{4}\sin^{2}(\theta)\left(\sin(\theta+\varphi)+1\right)\right]^{-1}
\eeq
\\
We now compute the term $\left(\Omega_{\scriptscriptstyle{E}}(x)\right)_{abc}$ :
\beq
\left(\Omega_{\scriptscriptstyle{E}}(x)\right)_{abc}\ &=& \vert \det e\, \vert_{\vert x}\ \sum_{i\, =\, 1}^{2}\ \alpha_{i}\
\left(^{\varphi_{i}\!}\omega(\varphi_{i}(x))\right)_{ab\mu}\
{\left(^{\varphi_{i}\!}e^{-1}(\varphi_{i}(x))\right)^{\mu}}_{c} \\
\nnum &=& \vert \det e\, \vert_{\vert x}\
\left[\left(^{Id\!}\omega(x)\right)_{ab\mu}\
{\left(^{Id\!}e^{-1}(x)\right)^{\mu}}_{c}\ +\
\left(^{\sigma^{-1}\!}\omega(\sigma^{-1}(x))\right)_{ab\mu}\
{\left(^{\sigma^{-1}\!}e^{-1}(\sigma^{-1}(x))\right)^{\mu}}_{c}\right] \\
\nnum &=& \vert \det e\, \vert_{\vert x}\
\left[\left(\omega(x)\right)_{ab\mu}\
{\left(e^{-1}(x)\right)^{\mu}}_{c}\ +\
\left(\tilde\omega(\sigma^{-1}(x))\right)_{ab\mu}\
{\left(\tilde e^{-1}(\sigma^{-1}(x))\right)^{\mu}}_{c}\right] \\
\nnum = r^{2}\sin(\theta) \sp\sp && \left[\left(\bar{ccccc} \left(0_{3}\right)_{(ab)} &,& \left(J_{z}\right)_{(ab)} &,& \left(\cos(\theta)J_{x}-\sin(\theta)J_{y}\right)_{(ab)} \ear\right)_{\!\!(\mu)}
\left(\bar{ccc}1&0&0\\0&\frac{1}{r}&0\\0&0&\frac{1}{r\sin(\theta)}\ear\right)_{\sp(\mu\,c)}\right] \\
\nnum &=& \left(\bar{ccccc} \left(0_{3}\right)_{(ab)} &,& r\sin(\theta)\left(J_{z}\right)_{(ab)} &,& r\left(\cos(\theta)J_{x}-\sin(\theta)J_{y}\right)_{(ab)} \ear\right)_{\!\!(c)}
\eeq
\\
Thanks to these terms, we express all elements composing the fluctuated Dirac operator. \\
We determine the expression of the fluctuated vierbein and its determinant:
\beq
{\left(\ef (x) \right)^{a}}_{\mu} &=& \vert \det \left(E (x) \right)\vert^{\frac{-1}{3-1}}\ {\left(E(x)\right)^{a}}_{\mu} \\
\nnum &&\sp\sp\sp\sp =
r^{2}\sin(\theta)\sqrt{2\left(\sin(\theta+\varphi)+1\right)}\ \left(\bar{ccc}
\frac{\csc(\theta)}{2r^{2}} & \frac{\csc(\theta)}{2r\left(\sec(\theta+\varphi)+\tan(\theta+\varphi)\right)} & -\frac{\cos(\theta)+\sin(\varphi)}{2r\left(\sin(\theta+\varphi)+1\right)} \\
-\frac{\csc(\theta)\cos(\theta+\varphi)}{2r^{2}\left(\sin(\theta+\varphi)+1\right)}  & \frac{\csc(\theta)}{2r} & \frac{\sin(\theta)+\cos(\varphi)}{2r\left(\sin(\theta+\varphi)+1\right)} \\
\frac{\csc(\theta)\left(\cos(\theta)+\sin(\varphi)\right)}{2r^{2}\left(\sin(\theta+\varphi)+1\right)} & -\frac{\csc(\theta)\left(\sin(\theta)+\cos(\varphi)\right)}{2r\left(\sin(\theta+\varphi)+1\right)} & \frac{1}{2r}
\ear\right)
\eeq
and
\beq
\det \ef (x) &=& \vert \det \left(E (x) \right)\vert^{\frac{-1}{3-1}} \\
\nnum &=& r^{2}\sin(\theta)\sqrt{2\left(\sin(\theta+\varphi)+1\right)}
\eeq
where the secant is defined by: $\sec(\theta) = 1/\cos(\theta)$.\\
Once more, we observe that we are no longer in the symmetric gauge. Now, the formula:
\beq
\left({^{f}\!g}(x)\right)_{\mu\nu} = \left(\ef^{T}\ \ef\right)_{\mu\nu}
\eeq
allows us to obtain the metric tensor corresponding to the fluctuated metric $^{f}\!g(x)$, expressed in the frame $\pa_{\mu}$ :
\beq
\\
\nnum \left({^{f}\!g}(x)\right)_{\mu\nu} = \left(\bar{ccc}
 \frac{3-\sin(\theta-\varphi)}{2} & \frac{-r\cos(\theta-\varphi)}{2} &
 \frac{r\sin(\theta)\left[-\cos(\theta)+\sin(\varphi)\right]}{2}\\
 \frac{-r\cos(\theta-\varphi)}{2} & \frac{3-\sin(\theta-\varphi)}{2}\ r^{2} &
 \frac{-r^{2}\sin(\theta)\cos(\theta+\varphi)\left[\cos(\theta)+\sin(\varphi)\right]}{2\left[\sin(\theta+\varphi)+1\right]}\\
 \frac{r\sin(\theta)\left[-\cos(\theta)+\sin(\varphi)\right]}{2} & \frac{-r^{2}\sin(\theta)\cos(\theta+\varphi)\left[\cos(\theta)+\sin(\varphi)\right]}{2\left[\sin(\theta+\varphi)+1\right]} &
 \frac{3-\sin(\theta-\varphi)}{2}\ r^{2}\sin^{2}(\theta)\ear\right)
\eeq

We can now calculate the last element composing the fluctuated Dirac operator, the fluctuated spin connection:
\beq
\nnum\left(\omegaf(x)\right)_{ab\mu}\!\!\! &=& \left(\Omega_{\scriptscriptstyle{E}}(x)\right)_{abc}\
{\left(E(x)\right)^{c}}_{\mu} \\
&=& \left(\bar{ccccc} \left(0_{3}\right)_{(ab)} &,& r\sin(\theta)\left(J_{z}\right)_{(ab)} &,&
r\left(\cos(\theta)J_{x}-\sin(\theta)J_{y}\right)_{(ab)} \ear\right)_{\!\!(c)} {}\\
\nnum &&{}\ \ \ \ \ \ \ \ \ \ \ \ \left(\bar{ccc}
\frac{\csc(\theta)}{2r^{2}} & \frac{\csc(\theta)}{2r\left(\sec(\theta+\varphi)+\tan(\theta+\varphi)\right)} & -\frac{\cos(\theta)+\sin(\varphi)}{2r\left(\sin(\theta+\varphi)+1\right)} \\
-\frac{\csc(\theta)\cos(\theta+\varphi)}{2r^{2}\left(\sin(\theta+\varphi)+1\right)}  & \frac{\csc(\theta)}{2r} & \frac{\sin(\theta)+\cos(\varphi)}{2r\left(\sin(\theta+\varphi)+1\right)} \\
\frac{\csc(\theta)\left(\cos(\theta)+\sin(\varphi)\right)}{2r^{2}\left(\sin(\theta+\varphi)+1\right)} & -\frac{\csc(\theta)\left(\sin(\theta)+\cos(\varphi)\right)}{2r\left(\sin(\theta+\varphi)+1\right)} & \frac{1}{2r}
\ear\right)_{\sp(c\mu)}
\eeq
Recalling that latin letters correspond to components expressed in the mobile frame, and that space is euclidian:
${\left(\omegaf(x)\right)^{a}}_{b\mu} = \delta^{ac}\ \left(\omegaf(x)\right)_{cb\mu}$.

\beq
\left\{\bar{l}
{\left(\omegaf(x)\right)^{a}}_{b\,1} = \left(\bar{ccc}
 0 & \frac{\cos(\theta+\varphi)}{2r\left(\sin(\theta+\varphi)+1\right)} &
 -\frac{\cos(\theta)+\sin(\varphi)}{2r\left(\sin(\theta+\varphi)+1\right)}\\
 -\frac{\cos(\theta+\varphi)}{2r\left(\sin(\theta+\varphi)+1\right)} & 0 &
 -\frac{\cot(\theta)\left(\cos(\theta)+\sin(\varphi)\right)}{2r\left(\sin(\theta+\varphi)+1\right)}\\
 \frac{\cos(\theta)+\sin(\varphi)}{2r\left(\sin(\theta+\varphi)+1\right)} & \frac{\cot(\theta)\left(\cos(\theta)+\sin(\varphi)\right)}{2r\left(\sin(\theta+\varphi)+1\right)} & 0
  \ear\right) \\
  \textrm{} \\
{\left(\omegaf(x)\right)^{a}}_{b\,2} = \left(\bar{ccc}
 0 & -\frac{1}{2} & \frac{\sin(\theta)+\cos(\varphi)}{2\left(\sin(\theta+\varphi)+1\right)} \\
 \frac{1}{2} & 0 &
 \frac{\cot(\theta)\left(\sin(\theta)+\cos(\varphi)\right)}{2\left(\sin(\theta+\varphi)+1\right)}\\
 -\frac{\sin(\theta)+\cos(\varphi)}{2\left(\sin(\theta+\varphi)+1\right)} & -\frac{\cot(\theta)\left(\sin(\theta)+\cos(\varphi)\right)}{2\left(\sin(\theta+\varphi)+1\right)} & 0
  \ear\right) \\
  \textrm{} \\
{\left(\omegaf(x)\right)^{a}}_{b\,3} = \left(\bar{ccc}
 0 & -\frac{\sin(\theta)\left(\sin(\theta)+\cos(\varphi)\right)}{2\left(\sin(\theta+\varphi)+1\right)} &
 -\frac{\sin(\theta)}{2}\\
 \frac{\sin(\theta)\left(\sin(\theta)+\cos(\varphi)\right)}{2\left(\sin(\theta+\varphi)+1\right)} & 0 &
 -\frac{\cos(\theta)}{2}\\
 \frac{\sin(\theta)}{2} & \frac{\cos(\theta)}{2} & 0
  \ear\right)
  \ear\right.
\eeq

At this step of the calculation, the use of Mathematica become necessary. We first verify the hermiticity of the operator. We previously showed that this requirement corresponds to the constraint:
\beq
{^{f}T^{\,a}}_{a\mu} = 0
\eeq
We calculate the torsion 2-form ${^{f}T(x)}$ :
\beq
{^{f}T}(x) =\ \ d\ef(x)\ +\ \omegaf(x)\wedge\ef(x)\ = \frac{1}{2}\ {^{f}T^{\,a}}_{\mu\nu}\ dx^{\mu}\wedge
dx^{\nu}
\eeq
The expression of the torsion term is rather hard to read, we report it in the appendix. This term has, at least, the usefulness to show that the torsion of the resulting space is non-vanishing. The torsion 2-form give, in components, the tensor ${^{f}T^{\,a}}_{\mu\nu}$. We thus have to express the first covariant index in the fluctuated mobile frame, by using the fluctuated vierbein:
\beq
{^{f}T^{\,a}}_{b\mu} = {^{f}T^{\,a}}_{\rho\mu} {\left(\ef^{-1}\right)^{\rho}}_{b}
\eeq
We finally obtain that the hermiticity condition is satisfied:
\beq
{^{f}T^{\,a}}_{a\mu} = {^{f}T^{\,a}}_{\rho\mu} {\left(\ef^{-1}\right)^{\rho}}_{a} = 0
\eeq
It still remains to show that this fluctuation has generated, in addition of a non-vanishing torsion, a non-vanishing curvature. To this aim, we have to compute the fluctuated curvature scalar. We begin with the calculation of the curvature 2-form ${^{f}\!R(x)}$:
\beq
{^{f}\!R(x)} =\ \ d\omegaf(x)\ +\ \omegaf(x)\wedge\omegaf(x)
\eeq
Here again, the explicit expression of the curvature 2-form is not relevant in itself, we report it in the appendix. The following steps necessary to the calculation of the curvature scalar, identical to the previous example, are briefly summarized here after:
\beq
{^{f}\!R} &=& \frac{1}{2}\ {{^{f}\!R}^{a}}_{b\mu\nu}\ dx^{\mu}\wedge dx^{\nu}\ \ \ \ \ \ \ \ \ \ \ \ \ \ \ \ \ \ \, \textrm{(2-forme de courbure)}\\
\nnum {{^{f}\!R}^{\alpha}}_{\beta\mu\nu} &=& {\left(\ef^{-1}\right)^{\alpha}}_{a}\ {{^{f}\!R}^{a}}_{b\mu\nu}\  {\ef^{b}}_{\beta}\ \ \ \ \ \ \ \ \ \ \ \ \ \ \ \ \ \textrm{(Riemann tensor)} \\
\nnum {{^{f}\!R}}_{\mu\nu} &=& {{^{f}\!R}^{\alpha}}_{\mu\alpha\nu}\ =\ {\left(\ef^{-1}\right)^{\alpha}}_{a}\ {{^{f}\!R}^{a}}_{b\alpha\nu}\  {\ef^{b}}_{\mu}\ \ \ \ \ \textrm{(Ricci tensor)} \\
\nnum {{^{f}\!R}}_{S} &=& {{^{f}\!R}}_{\mu\nu}\ {^{f}\!g}^{\mu\nu}\ \ \ \ \ \ \ \ \ \ \ \ \ \ \ \ \ \ \ \ \ \ \ \ \ \ \ \ \ \ \ \ \textrm{(scalaire de courbure)}
\eeq
We finally obtain the curvature scalar characterizing the space described by the fluctuated Dirac operator:
\beq
\nnum {{^{f}\!R}}_{S} &=& \frac{-\csc^{2}(\theta)}{8r^{2}\left[\cos\left(\frac{\theta+\varphi}{2}\right)+\sin\left(\frac{\theta+\varphi}{2}\right)\right]^{3}}\ \left[2\cos(\frac{5\theta-3\varphi}{2})+\cos(\frac{3\theta-\varphi}{2})-\cos(\frac{7\theta-\varphi}{2}) \right. {}\\
      &&{} \left.-21\cos(\frac{\theta+\varphi}{2})+4\cos(\frac{3(\theta+\varphi)}{2})+\cos(\frac{5\theta+\varphi}{2})-2\cos(\frac{7\theta+3\varphi}{2})\right. {}\\
\nnum &&{} \left.\hspace{5.5cm}+2\sin(\frac{5\theta-3\varphi}{2})-\sin(\frac{3\theta-\varphi}{2})+\sin(\frac{7\theta-\varphi}{2})\right.\\
\nnum &&{} \left.-21\sin(\frac{\theta+\varphi}{2})-4\sin(\frac{3(\theta+\varphi)}{2})+\sin(\frac{5\theta+\varphi}{2})+2\sin(\frac{7\theta+3\varphi}{2})\right]
\eeq
This second example allows to confirm that fluctuations of the Dirac operator generate, at the same time, torsion and curvature from an initially flat and without torsion space. Moreover, the computational complexity of this example should be seen as an evidence of the validity of this fluctuation method. \\

{\renewcommand{\thechapter}{}\renewcommand{\chaptername}{}
\addtocounter{chapter}{5}
\chapter{Conclusion and outlook}\markboth{\sl CONCLUSION}{\sl CONCLUSION}}

At the end of this report, we have shown that the fluctuation method developed during this `stage', in the case of a commutative spectral triple (algebraic equivalent of a Riemannian space) is consistent. Successfully applied on two non trivial examples (two diffeomorphisms of euclidian spaces of dimension two and three), this method provides an additional agreement to the conjecture from which we have been inspired.\\

The substantial contribution from these fluctuations is to generate, from an initially flat space, a space related to a fluctuated Dirac operator which possess non vanishing curvature and torsion. This allows one to obtain an exact equivalence principle.\\

However, all along this report, we have been restricted to operate the fluctuation method on Dirac operators describing initially flat spaces. This has been done knowingly. Indeed, so far, we are not able to explicitly compute the spin connections belonging to spaces with torsion. The formulae (\ref{Connection}), that we use in order to determine the connection, has come from the first Cartan structure equation in which we have states $T=0$:
\beq
\omega\wedge e = -\,de
\eeq
But the generation to spaces with torsion imposes to enlarge the dynamical configuration space $\ff$ to the corresponding Dirac operators, and thus requires being able to compute their spin connections. The term $\omega$ solution of the equation:
\beq
\omega\wedge e = T - de
\eeq
appears as an inevitable next step.\\
\vspace{1cm}

Beyond this ability, enlarging the dynamical configuration space to operators  describing spaces with torsion leads directly to the question about the role of torsion in Noncommutative Geometry. So far, this question has been ignored although the Chamseddine-Connes spectral action makes it appear explicitly.

\appendix

\chapter{Annexes}

\section{Notations}
\begin{center}
When not otherwise stated, the following notations are used in this report:
\end{center}

\beq\nnum&&\hspace{-1cm}\bar{ll}
M & \textrm{differentiable manifold of dimension $n$ standing for space-time.}\\
\uu & \textrm{open set of $M$ isomorphic to an open set of $\RR^{n}$} \\
\\
\beta_{\mu}(x) & \textrm{frame defined $\forall x \in \uu$} \\
\beta^{\mu}(x) & \textrm{dual frame} \\
e_{a}(x) & \textrm{orthonormal frame associated to the metric (vierbein)} \\
\\
g(x) & \textrm{metric defined on $M$} \\
\left(g(x)\right)_{\mu\nu} & \textrm{metric tensor (corresponding to a frame)} \\
\eta & \textrm{Minkowski metric describing a flat, torsionless space} \\
\eta_{\mu\nu} & \textrm{metric tensor associated to the Minkowski metric in an inertial frame} \\
\\
\textrm{Diff($M$)} & \textrm{Group of diffeomorphisms on $M$} \\
GL_{4}\left(\RR\right)& \textrm{General Linear group with real constants as entries} \\
^{\uu}GL_{4}\left(\RR\right) & \textrm{General Linear group with real functions on $\uu$ as entries} \\
SO\left(3,1\right) & \textrm{Special Orthogonal group} \\
^{\uu}SO\left(3,1\right) & \textrm{Special Orthogonal group with real functions on $\uu$ as entries} \\
Spin\left(3,1\right) & \textrm{Universal covering group of $SO\left(3,1\right)$} \\
\ear \\
\nnum &&\hspace{-1cm}\bar{lcl}
\varphi & \hspace{0.8cm}\textrm{Element of} & \textrm{Diff($M$)} \\
\Lambda & & \textrm{$GL_{4}\left(\RR\right)$} \\
\Lambda(x) & & \textrm{$^{\uu}GL_{4}\left(\RR\right)$} \\
\Lambda_{\varphi}(x) & & \textrm{$^{\uu}GL_{4}\left(\RR\right)$, which is the jacobian of $\varphi$} \\
\Lambda_{_{\ll}} & & \textrm{$SO\left(3,1\right)$} \\
\Lambda_{_{\ll}}(x) & & \textrm{$^{\uu}SO\left(3,1\right)$} \\
\ear\eeq

\newpage
\section{Some useful definitions and theorems}
{\bfseries Definition} :\ \ \  A frame $\beta_{\mu}(x)$ defined on an open set $\uu \subset \RR^{n}$ is an set of $n$ differentiable vectorial fields :
\beq
\beta_{\mu}(x) := \left\{\beta_{\mu}(x)\right\}_{\mu\,=\,1}^{n} =  \left\{\beta_{1}(x),..,\beta_{n}(x)\right\}
\eeq
such that for every point $x \in \uu$, $\left\{\beta_{\mu\,\vert x}\right\}_{\mu\,=\,1}^{n}$ is a basis.\\
In other words, a frame is a differentiable field of bases.\\

{\bfseries Definition} :\ \ \  The dual frame $\beta^{\mu}(x)$ of a frame $\beta_{\nu}(x)$ is defined by the relation:
\beq
\hspace{3cm}\beta^{\mu}(x)\ \beta_{\nu}(x) = \delta^{\mu}_{\nu}\hspace{2cm} \forall x \in \uu
\eeq

{\bfseries Definition} :\ \ \  A frame $\beta_{\mu}(x)$ is said orthonormal for a metric $g$, if the metric tensor $\left(g(x)\right)_{\mu\nu}$ in this frame is such that:
\beq
\hspace{3cm}\left(g(x)\right)_{\mu\nu} = g\left(\beta_{\mu},\beta_{\nu}\right)_{\vert x} = \eta_{\mu\nu}\hspace{2cm} \forall x \in \uu
\eeq

{\bfseries Definition} :\ \ \  A frame $\beta^{\mu}(x)$ is said holonomic (or deriving from a coordinate system) if it is such that:
\beq
\hspace{3cm}d\beta^{\mu} &=& 0\hspace{2cm} \forall \mu\,=\,1,..,n
\eeq

{\bfseries Theorem ({\small \cite{SCH01}})} :\ \ \  A subset $\uu$ of $\RR^{n}$ admits a holonomic and orthonormal frame if and only if this subset is flat. \\

{\bfseries Theorem ({\small Gram-Schmidt})} :\ \ \ Every metric $g$ admits an orthonormal basis, i.e. a basis $\left\{e_{1},...,e_{n}\right\}$ such that:
\vspace{-0.2cm}
\beq
g\left(e_{a},e_{b}\right) = \eta_{ab}
\eeq
with $\eta_{ab}$ being the minkowskian metric tensor of the metric $g$ in the basis $\left\{e_{a}\right\}_{a\,=\,1}^{n}$.\\
\indent The Gram-Schmidt orthonormalization algorithm allows one, starting from an arbitrary basis on a point $x \in \uu$, to get an orthonormal basis.
This algorithm requires only addition, multiplication and division operations, which preserve differentiability. Hence, it ensures that, starting from an arbitrary frame $\beta_{\mu}$, one can construct an orthonormal frame $e_{a}$. \\

\newpage
\section{Details of some calculations}

\subsection{Invariant volume form}
We will explicitly show the invariance of the volume form under diffeomorphisms. We will then show that this property, together with the theorem of change of variables, allows one to express the Dirac action integral under a change of coordinates. For simplicity, this calculation takes place in a 4-dimensional space, but the generalization to an arbitrary finite dimension is straightforward. \\

\noi We consider a diffeomorphism $\varphi$ linking two coordinates system ${\tilde x}^{\mu}$ and $x^{\mu}$, with their respective frames $\{\tilde\pa_{\mu}\}_{\mu\, =\, 0}^{3}$ and $\{\pa_{\mu}\}_{\mu\, =\, 0}^{3}$, such that:
\beq
    x^{\mu} = \varphi^{\mu}(\tilde x)
\eeq
We have the following relations:
\beq
\label{Conj001}dx^{\mu} &=& \frac{\pa \varphi^{\mu}(\tilde x)}{\pa \tilde x^{\nu}}\ d\tilde x^{\nu} =: {\left(J(\tilde x)\right)^{\mu}}_{\nu}\ d\tilde x^{\nu} \\
\nnum d\tilde x^{\mu} &=& {\left(J^{-1}(\tilde x)\right)^{\mu}}_{\nu}\ dx^{\nu} = {\left(J^{-1}\left(\varphi^{-1}(x)\right)\right)^{\mu}}_{\nu}\ dx^{\nu} \\
\nnum {\left(J(\tilde x)\right)^{\mu}}_{\nu} &=& {\left(J\left(\varphi^{-1}(x)\right)\right)^{\mu}}_{\nu}
\eeq
where $J$ is the jacobian of the diffeomorphism $\varphi$ and $J^{-1}$ its inverse. Each frame is deduced from the other by applying the jacobian or its inverse such that:
\beq
\pa_{\mu} &=& {\left(J^{-1}(\tilde x)\right)^{\nu}}_{\mu}\ \tilde\pa_{\nu} \\
\nnum \tilde\pa_{\mu} &=& {\left(J(\tilde x)\right)^{\nu}}_{\mu}\ \pa_{\nu}
\eeq

\noi In the frame $\tilde\pa_{\mu}$, the invariant volume form $d\tilde V_{4}$ is expressed as:
\beq
\label{Conj002}d\tilde V_{4} = \sqrt{\vert\det\tilde g_{..}\vert}_{\vert\tilde x}\ d^{4}\tilde x\ =\ \sqrt{\vert\det\tilde g_{..}\vert}_{\vert\tilde x}\ d\tilde x^{0}\wedge d\tilde x^{1}\wedge d\tilde x^{2}\wedge d\tilde x^{3}
\eeq
where $\tilde g_{..}$ is the metric tensor associated to the frame $\tilde\pa_{\mu}$. \\
One can connect the metric tensors $\tilde g_{\mu\nu}$ and $g_{\mu\nu}$, respectively belonging to $\tilde\pa_{\mu}$ and $\pa_{\mu}$, by expressing the metric in each frame:
\beq
\label{Conj004} g &=& \tilde g(\tilde x)_{\mu\nu}\ d\tilde x^{\mu}\otimes d\tilde x^{\nu} \\
\nnum &=& g(x)_{\rho\sigma}\ dx^{\rho}\otimes dx^{\sigma} \\
\nnum &=& g(x)_{\rho\sigma}\ {\left(J(\tilde x)\right)^{\rho}}_{\mu}\ {\left(J(\tilde x)\right)^{\sigma}}_{\nu}\ d\tilde x^{\mu}\otimes d\tilde x^{\nu} \\
\nnum \Longrightarrow\ \ \tilde g(\tilde x)_{\mu\nu} &=& {(J^{T}(\tilde x))_{\mu}}^{\rho}\ g\left(\varphi(\tilde x)\right)_{\rho\sigma}\ {\left(J(\tilde x)\right)^{\sigma}}_{\nu}
\eeq
Equation (\ref{Conj002}) becomes :
\beq
\label{Conj003}\sqrt{\vert\det\tilde g_{..}\vert}_{\vert\tilde x}\ d^{4}\tilde x\ &=& \sqrt{\left\vert\det\left[\left(J^{T}_{\vert\tilde x}\ g_{\vert\varphi(\tilde x)} J_{\vert\tilde x}\right)_{..}\ \right]\right\vert}\ d\tilde x^{0}\wedge d\tilde x^{1}\wedge d\tilde x^{2}\wedge d\tilde x^{3} \\
\nnum &=& \vert\det J\vert_{\vert\tilde x}\ \sqrt{\vert\det g_{..}\vert}_{\vert\varphi(\tilde x)}\ d\tilde x^{0}\wedge d\tilde x^{1}\wedge d\tilde x^{2}\wedge d\tilde x^{3}
\eeq
Moreover, the relations (\ref{Conj001}) lead to the equality :
\beq
d\tilde x^{0}\wedge d\tilde x^{1}\wedge d\tilde x^{2}\wedge d\tilde x^{3} &=& {(J^{-1}_{\vert\tilde x})^{0}}_{\mu}\ {(J^{-1}_{\vert\tilde x})^{1}}_{\nu}\ {(J^{-1}_{\vert\tilde x})^{2}}_{\rho}\ {(J^{-1}_{\vert\tilde x})^{3}}_{\sigma}\ dx^{\mu}\wedge dx^{\nu}\wedge dx^{\rho}\wedge dx^{\sigma} \\
\nnum &=& {(J^{-1}_{\vert\tilde x})^{0}}_{\mu}\ {(J^{-1}_{\vert\tilde x})^{1}}_{\nu}\ {(J^{-1}_{\vert\tilde x})^{2}}_{\rho}\ {(J^{-1}_{\vert\tilde x})^{3}}_{\sigma}\ {\epsilon}^{\,\mu\nu\rho\sigma}\ dx^{0}\wedge dx^{1}\wedge dx^{2}\wedge dx^{3} \\
\nnum &=& \det\left(J^{-1}_{\vert\tilde x}\right)\ dx^{0}\wedge dx^{1}\wedge dx^{2}\wedge dx^{3} \\
\nnum &=& \left(\det J\right)^{-1}_{\vert\tilde x}\ d^{4}x
\eeq
where ${\epsilon}^{\,\mu\nu\rho\sigma}$, the skew-symmetric Lévy-Civita tensor, is obtained from the permutation of the indices of the volume form. We then use the general expression of the determinant in order to identify the volume form \cite{CHO01}. \\
One finally obtains:
\beq
\sqrt{\vert\det\tilde g_{..}\vert}_{\vert\tilde x}\ d^{4}\tilde x\ &=& \vert\det J\vert_{\vert\tilde x}\ \sqrt{\vert\det g_{..}\vert}_{\vert\varphi(\tilde x)}\ \left(\det J\right)^{-1}_{\vert\tilde x}\ d^{4}x \\
\nnum &=& \vert\det J\vert_{\vert\varphi^{-1}(x)}\ \sqrt{\vert\det g_{..}\vert}_{\vert\varphi(\varphi^{-1}(x))}\ \left(\det J\right)^{-1}_{\vert\varphi^{-1}(x)}\ d^{4}x \\
\nnum &=& \sqrt{\vert\det g_{..}\vert}_{\vert\varphi(\varphi^{-1}(x))}\ d^{4}x \\
\nnum &=& \sqrt{\vert\det g_{..}\vert}_{\vert x}\ d^{4}x
\eeq
where the equality of the third line is obtained by considering only the diffeomorphisms $\varphi$ for which the determinant of the jacobian is positive. \\
Hence, we have shown that:
\beq
d\tilde V_{4} = \sqrt{\vert\det\tilde g_{..}\vert}_{\vert\tilde x}\ d^{4}\tilde x = \sqrt{\vert\det g_{..}\vert}_{\vert x}\ d^{4}x = dV_{4}
\eeq
for a diffeomorphism $\varphi$ such that $x = \varphi(\tilde x)$ preserving orientation (i.e. $\det J > 0$). \\

\newpage
\subsection{Mathematica code of example 2}
\begin{scriptsize}
\begin{verbatim}
**** Introduction of the inverse jacobian Ji :
Ji = {{Sin[\[Theta]]Cos[\[Phi]], r Cos[\[Theta]]Cos[\[Phi]], -r Sin[\[Theta]]Sin[\[Phi]]},
     {Sin[\[Theta]]Sin[\[Phi]], r Cos[\[Theta]]Sin[\[Phi]], r Sin[\[Theta]]Cos[\[Phi]]},
     {Cos[\[Theta]],-r Sin[\[Theta]], 0}}; MatrixForm[Ji]
J = FullSimplify[Inverse[Ji]]; MatrixForm[J]

**** Introduction of vierbein e :
e = {{1, 0, 0}, {0, r, 0}, {0, 0, r Sin[\[Theta]]}}; MatrixForm[e]
ei = FullSimplify[Inverse[e]]; MatrixForm[ei]

**** Computation of determinants :
detJi = FullSimplify[Det[Ji]]
dete = FullSimplify[Det[e]]

**** Introduction of the spin connection \[omega] :
J0 = {{0, 0, 0}, {0, 0, 0}, {0, 0, 0}}; MatrixForm[J0]
Jx = {{0, 0, 0}, {0, 0, -1}, {0, 1, 0}}; MatrixForm[Jx];
Jy = {{0, 0, 1}, {0, 0, 0}, {-1, 0, 0}}; MatrixForm[Jy];
Jz = {{0, -1, 0}, {1, 0, 0}, {0, 0, 0}}; MatrixForm[Jz]
Jxy = Cos[\[Theta]]Jx - Sin[\[Theta]] Jy; MatrixForm[Jxy]
\[Omega]tilde = {J0, Jz, Jxy}; MatrixForm[\[Omega]tilde];
\[Omega] = Transpose[\[Omega]tilde, {3, 1, 2}]; MatrixForm[\[Omega]];
MatrixForm[\[Omega][[All, All, 1]]]
MatrixForm[\[Omega][[All, All, 2]]]
MatrixForm[\[Omega][[All, All, 3]]]

**** Computation of term Ei and its determinant:
Ei = FullSimplify[detJi J + dete ei]; MatrixForm[Ei]
detEi = FullSimplify[Det[Ei]];
detef = detEi^(1/2)

**** Computation of the fuctuated vierbein:
efi = Ei/detef; MatrixForm[efi]
ef = FullSimplify[Inverse[efi]]; MatrixForm[ef]

**** Computation of the fluctuated spin connection:
\[Omega]f = dete \[Omega].ei.ef/detef; MatrixForm[\[Omega]f];
MatrixForm[\[Omega]f[[All, All, 1]]]
MatrixForm[\[Omega]f[[All, All, 2]]]
MatrixForm[\[Omega]f[[All, All, 3]]]

**** Computation of the fluctuated torsion 2-form:
def12 = FullSimplify[D[ef[[All, 2]], r] - D[ef[[All, 1]], \[Theta]]]; MatrixForm[def12]
def23 = FullSimplify[D[ef[[All, 3]], \[Theta]] - D[ef[[All, 2]], \[Phi]]]; MatrixForm[def23]
def31 = FullSimplify[D[ef[[All, 1]], \[Phi]] - D[ef[[All, 3]], r]]; MatrixForm[def31]
def00 = {0, 0, 0}; MatrixForm[def00];
deftilde = {{def00, -def12, def31}, {def12, def00, -def23}, {-def31, def23, def00}};
def = Transpose[deftilde, {3, 2, 1}]; MatrixForm[def];
MatrixForm[def12] === MatrixForm[def[[All, 1, 2]]];
MatrixForm[def12] === MatrixForm[-def[[All, 2, 1]]];
MatrixForm[def23] === MatrixForm[def[[All, 2, 3]]];
MatrixForm[def23] === MatrixForm[-def[[All, 3, 2]]];
MatrixForm[def31] === MatrixForm[def[[All, 3, 1]]];
MatrixForm[def31] === MatrixForm[-def[[All, 1, 3]]];

\[Omega]fef12 = \[Omega]f[[All, All, 1]].ef[[All, 2]] - \[Omega]f[[All, All, 2]].ef[[All, 1]]; MatrixForm[\[Omega]fef12]
\[Omega]fef23 = \[Omega]f[[All, All, 2]].ef[[All, 3]] - \[Omega]f[[All, All, 3]].ef[[All, 2]]; MatrixForm[\[Omega]fef23]
\[Omega]fef31 = \[Omega]f[[All, All, 3]].ef[[All, 1]] - \[Omega]f[[All, All, 1]].ef[[All, 3]]; MatrixForm[\[Omega]fef31]
\[Omega]feftilde = {{def00, -\[Omega]fef12, \[Omega]fef31}, {\[Omega]fef12, def00, -\[Omega]fef23},
                   {-\[Omega]fef31, \[Omega]fef23, def00}};
\[Omega]fef = Transpose[\[Omega]feftilde, {3, 2, 1}]; MatrixForm[\[Omega]fef];
MatrixForm[\[Omega]fef12] === MatrixForm[\[Omega]fef[[All, 1, 2]]];
MatrixForm[\[Omega]fef12] === MatrixForm[-\[Omega]fef[[All, 2, 1]]];
MatrixForm[\[Omega]fef23] === MatrixForm[\[Omega]fef[[All, 2, 3]]];
MatrixForm[\[Omega]fef23] === MatrixForm[-\[Omega]fef[[All, 3, 2]]];
MatrixForm[\[Omega]fef31] === MatrixForm[\[Omega]fef[[All, 3, 1]]];
MatrixForm[\[Omega]fef31] === MatrixForm[-\[Omega]fef[[All, 1, 3]]];

Tfa\[Mu]\[Nu] = 2(def + \[Omega]fef); MatrixForm[Tfa\[Mu]\[Nu]];
Tfa\[Nu]\[Mu] = Transpose[Tfa\[Mu]\[Nu], {1, 3, 2}]; MatrixForm[Tfa\[Nu]\[Mu]];
Tfa\[Mu]\[Nu][[All, 1, 2]] == -Tfa\[Nu]\[Mu][[All, 1, 2]];
Tfa\[Mu]\[Nu][[All, 2, 3]] == -Tfa\[Nu]\[Mu][[All, 2, 3]];
Tfa\[Mu]\[Nu][[All, 3, 1]] == -Tfa\[Nu]\[Mu][[All, 3, 1]];
Tfa\[Nu]b = Tfa\[Nu]\[Mu].efi; MatrixForm[Tfa\[Nu]b];
Tfab\[Nu] = Transpose[Tfa\[Nu]b, {1, 3, 2}]; MatrixForm[Tfab\[Nu][[All, All, 1]]];

**** Verification of the hermiticity of the fluctuated Dirac operator:
Tfaa1 = Tr[Tfab\[Nu][[All, All, 1]]];
FullSimplify[Tfaa1];
% === 0
True
Tfaa2 = Tr[Tfab\[Nu][[All, All, 2]]];
FullSimplify[Tfaa2];
% === 0
True
Tfaa3 = Tr[Tfab\[Nu][[All, All, 3]]];
FullSimplify[Tfaa3];
% === 0
True

**** Determination of the curvature 2-form:
d\[Omega]f12 = FullSimplify[D[\[Omega]f[[All, All, 2]], r] - D[\[Omega]f[[All, All, 1]], \[Theta]]];
    MatrixForm[d\[Omega]f12]
d\[Omega]f23 = FullSimplify[D[\[Omega]f[[All, All, 3]], \[Theta]] - D[\[Omega]f[[All, All, 2]], \[Phi]]];
    MatrixForm[d\[Omega]f23]
d\[Omega]f31 = FullSimplify[D[\[Omega]f[[All, All, 1]], \[Phi]] - D[\[Omega]f[[All, All, 3]], r]];
    MatrixForm[d\[Omega]f31]
d\[Omega]f00 = J0;
d\[Omega]ftilde = {{d\[Omega]f00, -d\[Omega]f12, d\[Omega]f31}, {d\[Omega]f12, d\[Omega]f00, -d\[Omega]f23},
                  {-d\[Omega]f31, d\[Omega]f23, d\[Omega]f00}};
d\[Omega]f = Transpose[d\[Omega]ftilde, {4, 3, 1, 2}]; MatrixForm[d\[Omega]f];
MatrixForm[d\[Omega]f12] === MatrixForm[d\[Omega]f[[All, All, 1, 2]]];
MatrixForm[d\[Omega]f12] === MatrixForm[-d\[Omega]f[[All, All, 2, 1]]];
MatrixForm[d\[Omega]f23] === MatrixForm[d\[Omega]f[[All, All, 2, 3]]];
MatrixForm[d\[Omega]f23] === MatrixForm[-d\[Omega]f[[All, All, 3, 2]]];
MatrixForm[d\[Omega]f31] === MatrixForm[d\[Omega]f[[All, All, 3, 1]]];
MatrixForm[d\[Omega]f31] === MatrixForm[-d\[Omega]f[[All, All, 1, 3]]];

\[Omega]f\[Omega]f12 = FullSimplify[\[Omega]f[[All, All, 1]].\[Omega]f[[All, All, 2]]
                                    - \[Omega]f[[All, All, 2]].\[Omega]f[[All, All,1]]];
    MatrixForm[\[Omega]f\[Omega]f12]
\[Omega]f\[Omega]f23 = FullSimplify[\[Omega]f[[All, All, 2]].\[Omega]f[[All, All, 3]]
                                - \[Omega]f[[All, All, 3]].\[Omega]f[[All, All, 2]]];
    MatrixForm[\[Omega]f\[Omega]f23]
\[Omega]f\[Omega]f31 = FullSimplify[\[Omega]f[[All, All, 3]].\[Omega]f[[All, All, 1]]
                                - \[Omega]f[[All, All, 1]].\[Omega]f[[All, All, 3]]];
    MatrixForm[\[Omega]f\[Omega]f31]
\[Omega]f\[Omega]ftilde = {{d\[Omega]f00, -\[Omega]f\[Omega]f12, \[Omega]f\ \[Omega]f31},
                          {\[Omega]f\[Omega]f12, d\[Omega]f00, -\[Omega]f\[Omega]f23},
                          {-\[Omega]f\[Omega]f31, \[Omega]f\[Omega]f23, d\[Omega]f00}};
\[Omega]f\[Omega]f = Transpose[\[Omega]f\[Omega]ftilde, {4, 3, 1, 2}]; MatrixForm[\[Omega]f\[Omega]f];
MatrixForm[\[Omega]f\[Omega]f12] === MatrixForm[\[Omega]f\[Omega]f[[All, All, 1, 2]]];
MatrixForm[\[Omega]f\[Omega]f12] === MatrixForm[-\[Omega]f\[Omega]f[[All, All, 2, 1]]];
MatrixForm[\[Omega]f\[Omega]f23] === MatrixForm[\[Omega]f\[Omega]f[[All, All, 2, 3]]];
MatrixForm[\[Omega]f\[Omega]f23] === MatrixForm[-\[Omega]f\[Omega]f[[All, All, 3, 2]]];
MatrixForm[\[Omega]f\[Omega]f31] === MatrixForm[\[Omega]f\[Omega]f[[All, All, 3, 1]]];
MatrixForm[\[Omega]f\[Omega]f31] === MatrixForm[-\[Omega]f\[Omega]f[[All, All, 1, 3]]];
Rfab\[Mu]\[Nu] = 2(d\[Omega]f + \[Omega]f\[Omega]f); MatrixForm[Rfab\[Mu]\[Nu]];

**** Determination of the curvature scalar:
Rfa\[Nu]\[Mu]b = Transpose[Rfab\[Mu]\[Nu], {1, 4, 3, 2}];
Rf\[Alpha]\[Nu]\[Mu]\[Beta] = FullSimplify[efi.Rfa\[Nu]\[Mu]b.ef];
RiemannF = Rf\[Alpha]\[Beta]\[Mu]\[Nu] = Transpose[Rf\[Alpha]\[Nu]\[Mu]\[Beta], {1, 4, 3, 2}];
Rf\[Alpha]\[Mu]\[Beta]\[Nu] = Transpose[Rf\[Alpha]\[Beta]\[Mu]\[Nu], {1, 3, 2, 4}];
RicciF = Rf\[Mu]\[Nu] = FullSimplify[Tr[Rf\[Alpha]\[Mu]\[Beta]\[Nu], Plus, 2]]; MatrixForm[Rf\[Mu]\[Nu]]
gfi = efi.Transpose[efi]; MatrixForm[gfi];
Rfup\[Mu]\[Nu] = gfi.Rf\[Mu]\[Nu];
ScalaireF = FullSimplify[Tr[Rfup\[Mu]\[Nu]]]
\end{verbatim}
\end{scriptsize}

\newpage
\subsection{Curvature and torsion obtained in example 2}

\noi We give here after the expression of the fluctuated curvature and torsion 2-forms related to example 2: \\

\noi $\bullet$ Fluctuated torsion 2-form: $T = \frac{1}{2}\ {^{f}T^{a}}_{\mu\nu}\ dx^{\mu}\wedge dx^{\nu}$

\beq
{^{f}T^{a}}_{1\,2} = \left(\bar{c} \frac{\cos(\theta+\varphi)}{\sqrt{2}\sqrt{\left(\sin(\theta +
\varphi)+1\right)}} \\ \frac{-\cos^{2}(\theta+\varphi)}{\sqrt{2}\left(\sin(\theta + \varphi)+1\right)^{3/2}} \\
\frac{\cos(2\theta)+2\sin(\theta)\cos(\varphi)+4\cos(\theta)\sin(\varphi)-1}{2\sin(\theta)\sqrt{2}\sqrt{\left(\sin(\theta
+ \varphi)+1\right)}}\ear\right)
\eeq
\beq
{^{f}T^{a}}_{2\,3} = \left(\bar{c} \frac{-r
\left(3\cos(\frac{3\theta-\varphi}{2})-\cos(\frac{\theta+\varphi}{2})-3\sin(\frac{3\theta-\varphi}{2})-\sin(\frac{\theta+\varphi}{2})\right)\left(\cos(\frac{\theta+\varphi}{2})+\sin(\frac{\theta+\varphi}{2})\right)^{3}}{2\sqrt{2}\left(\sin(\theta
+ \varphi)+1\right)^{3/2}} \\ \frac{r
\left(\cos(\theta)\left(\sin(\theta)+\cos(\varphi)\right)+2\sin(\theta)\sin(\varphi)\right)}{\sqrt{2}\sqrt{\left(\sin(\theta
+ \varphi)+1\right)}} \\ \frac{-r\cos(\theta)\cos^{2}(\theta+\varphi)}{\sqrt{2}\left(\sin(\theta +
\varphi)+1\right)^{3/2}} \ear\right)
\eeq
\beq
{^{f}T^{a}}_{3\,1} = \left(\bar{c}
\frac{\cos(\theta-\varphi)+\sin(2\theta)}{\sqrt{2}\sqrt{\left(\sin(\theta + \varphi)+1\right)}} \\
\frac{\left(\cos(\theta)+\sin(\varphi)\right)^{2}}{\sqrt{2}\left(\sin(\theta + \varphi)+1\right)^{3/2}} \\
\frac{\sin(\theta)-\cos(\varphi)}{\sqrt{2}\sqrt{\left(\sin(\theta + \varphi)+1\right)}} \ear\right)
\eeq
\\

\noi $\bullet$ Fluctuated curvature 2-form: $R = \frac{1}{2}\ {^{f}R^{a}}_{b\,\mu\nu}\ dx^{\mu}\wedge dx^{\nu}$

\beq
{^{f}R^{a}}_{b\,1\,2} = \left(\bar{ccc} 0 & A & -B \\
-A & 0 & -C \\
B & C & 0 \ear \right)\
\ \ \textrm{où}\ \ \left\{\bar{lc} A = & \frac{1}{r\left(\sin(\theta+\varphi)+1\right)} \\ B = &
\frac{\csc(\theta)}{r\left(\csc(\theta+\varphi)+1\right)} \\ C = & \frac{\csc(\theta)\cot(\theta)}{r}\ear\right.
\eeq
\beq
\\
\nnum{^{f}R^{a}}_{b\,2\,3} = \left(\bar{ccc} 0 & -D & E \\
D & 0 & -F \\
- E & F & 0
\ear \right)\ \ \textrm{où}\ \ \left\{\bar{ll} D = & \frac{2\cos(\theta-\varphi)+\sin(2\theta)}{2\left(\sin(\theta+\varphi)+1\right)} \\ E = &
\frac{\cos(\theta)}{2}\left(1-\frac{2}{\csc(\theta+\varphi)+1}+\frac{\left(\sin(\theta)+\cos(\varphi)\right)^{2}}{\left(\sin(\theta+\varphi)+1\right)^{2}}\right) \\ F = & \frac{-2\cos(\theta)\cot(\theta)+\sin(2\theta)\sin(\varphi)}{2\left(\sin(\theta+\varphi)+1\right)}\ear\right.
\eeq
\beq
{^{f}R^{a}}_{b\,3\,1} = \left(\bar{ccc} 0 & -G & H \\
G & 0 & -I \\
-H & I & 0 \ear \right)\
\ \ \textrm{où}\ \ \left\{\bar{lc} G = & \frac{1}{r\left(\sin(\theta+\varphi)+1\right)} \\ H = &
\frac{\cos^{2}(\theta)\cos(\varphi)-\sin(\theta)}{r\left(\sin(\theta+\varphi)+1\right)}  \\ I = &
\frac{\cos(\theta)\left(\sin(\theta)\cos(\varphi)+1\right)}{r\left(\sin(\theta+\varphi)+1\right)}\ear\right.
\eeq


\begin{thebibliography}{2}
\bibitem[BLA01]{BLA01} \textbf{BLAGOJEVIC M.}, \textit{Gravitation and gauge symmetries}, Institute of Physics Publishing, 2002
\bibitem[CHO01]{CHO01} \textbf{CHOQUET-BRUHAT Y.} \& \textbf{DEWITT-MORETTE C.}, \textit{Analysis, manifolds and physics}, NORTH-HOLLAND, 1982
\bibitem[FRI01]{FRI01} \textbf{FRIOT S.}, \textit{Géométrie commutative : unification Maxwell-Einstein}, Rapport de stage DEA au CPT, 2000
\bibitem[G$\&$S01]{GetS01} \textbf{GÖCKELER M.} \& \textbf{SCHÜCKER T.}, \textit{Differential geometry, gauge theories, and gravity}, CAMBRIDGE University Press, 1987
\bibitem[IOC01]{IOC01} \textbf{IOCHUM B.}, \textbf{SCHÜCKER T.}, \textit{Diffeomorphisms and orthonormal frames}, hep-th/0406213v2, 2004
\bibitem[KRA01]{KRA01} \textbf{KRAJEWSKI T.}, \textit{Géométrie Non Commutative et interactions fondamentales}, math-ph/9903047, 1999
\bibitem[SCH01]{SCH01} \textbf{SCHÜCKER T.}, \textit{Geometries and forces}, hep-th/9712095v1, 1997
\bibitem[SCH02]{SCH02} \textbf{SCHÜCKER T.}, \textit{Forces from Connes' geometry}, hep-th/0111236v2, 2001
\bibitem[SCH03]{SCH03} \textbf{SCHÜCKER T.}, \textit{Spin group and almost commutative geometry}, hep-th/0007047v1, 2000
\bibitem[STE01]{STE01} \textbf{STEPHAN C.}, \textit{Noncommutative Geometry and the standard model of particle physics}, Thèse au Christian-Albrechts-Universität de Kiel et CPT Marseille, 2005
\end{thebibliography}
\end{document}